\documentclass[nofootinbib,a4paper,12pt]{article}
\usepackage{jheppub}
\usepackage[T1]{fontenc}
\usepackage[utf8]{inputenc}
\usepackage{float}
\usepackage{slashed}
\usepackage{multicol,multirow}
\usepackage{amssymb}
\usepackage{amsmath}
\usepackage{subcaption}
\usepackage{array}
\usepackage{color}
\usepackage{hyperref}
\hypersetup{colorlinks=true}
\usepackage{longtable}
\usepackage{graphics,graphicx,epsfig,ulem,booktabs}
\numberwithin{equation}{section}

\definecolor{blue}{rgb}{0.2, 0.4, 1.0}
\definecolor{green}{rgb}{0.1,0.8,0.2}
\definecolor{orange}{rgb}{1.0,0.5,0.0}
\definecolor{cyan}{rgb}{0.0,0.75,0.8}

\newcolumntype{C}[1]{>{\centering\let\newline\\\arraybackslash\hspace{0pt}}m{#1}}

\newcommand{\MEs}{\langle O_1^q \rangle}
\newcommand{\MEt}{\langle O_2^q \rangle}

\newcommand{\GeV}{\, {\rm GeV}}
\newcommand{\GeVsq}{\, {\rm GeV}^2}

\title{Non-factorisable effects in the decays
\boldmath $\bar B_{s}^0 \to D_{s}^+ \pi^-$ and $\bar B^0 \to D^+ K^-$ from LCSR}

\preprint{SI-HEP-2023-15}

\author[a]{Maria Laura Piscopo,}
\author[a]{Aleksey V. Rusov}

\affiliation[a]{Physik Department, Universit\"{a}t Siegen, 
Walter-Flex-Str. 3, 57068 Siegen, Germany}

\emailAdd{maria.piscopo@uni-siegen.de}
\emailAdd{rusov@physik.uni-siegen.de}

\abstract{In light of the current discrepancies between the recent predictions based on
QCD factorisation (QCDF) and the experimental data for several non-leptonic colour-allowed two-body $B$-meson decays, we obtain new determinations of the non-factorisable soft-gluon contribution to the decays $\bar B_{s}^0 \to D_{s}^+ \pi^-$ and $\bar B^0 \to D^+ K^-$, using the framework of light-cone sum rule (LCSR), with a suitable three-point correlation function and $B$-meson light-cone distribution amplitudes. In particular, we discuss the problem associated with a double light-cone (LC) expansion of the correlator, and motivate future determinations of the three-particle $B$-meson matrix element with the gluon and the spectator quark aligned along different light-cone directions.
Performing a LC-local operator product expansion of the correlation function, we find, for both modes considered, the non-factorisable part of the amplitude to be sizeable and positive, however, with very large systematic uncertainties.
Furthermore, we also determine for the first time, using LCSR, the factorisable amplitudes at LO-QCD, and thus the corresponding branching fractions. Our predictions are in agreement with the experimental data and consistent with the results based on QCDF, although again within very large uncertainties. 
In this respect, we provide a rich outlook for future improvements and investigations. 
}

\begin{document}

\maketitle

\section{Introduction}
\begin{figure}[t]
    \centering
    \includegraphics[scale=0.45]{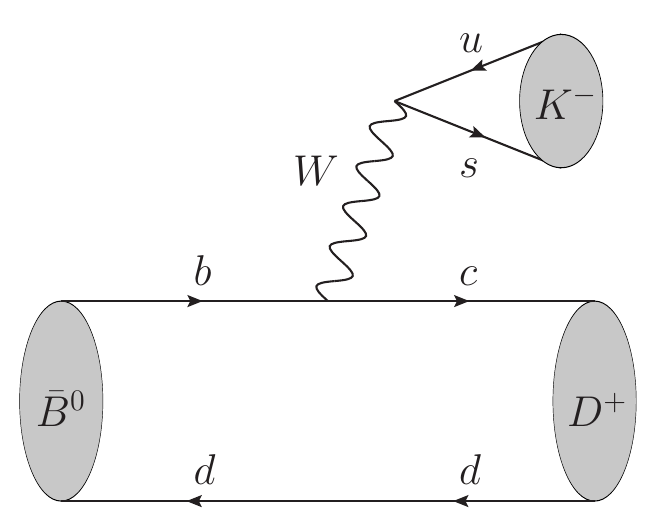}\qquad
    \includegraphics[scale=0.45]{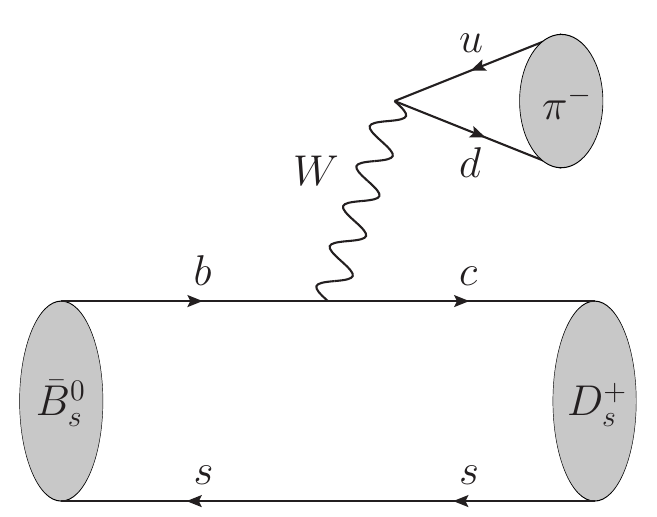}
    \caption{Schematic representation of the decays $\bar B^0 \to D^+ K^-$ and 
$\bar B_{s}^0 \to D_{s}^+ \pi^-$.}
    \label{fig:B-to-D-L-diagrams}
\end{figure}

The study of $B$-meson decays plays a crucial role in testing the Standard Model~(SM), as well as in searching or constraining possible New Physics~(NP) scenarios. 
Among these decays,  those involving only non-leptonic final states are notoriously the most challenging to be described, due to the complicated underlying hadronic structure. 
Their investigation, however, can allow one to test the different QCD based methods designed specifically for the study of these processes. 
In the present work, we focus on two particularly interesting examples of non-leptonic two-body $B$-meson decays, namely $\bar B^0 \to D^+ K^-$ and 
$\bar B_{s}^0 \to D_{s}^+ \pi^-$. In fact, as the flavour of all the quarks in the final state is different, see Figure~\ref{fig:B-to-D-L-diagrams}, neither the penguin nor the weak-annihilation topologies can contribute, and these are generally considered to be among the theoretically cleanest non-leptonic $B$-meson decays. 
By now these modes are determined experimentally quite precisely, and
the Particle Data Group (PDG) quotes the following values \cite{Workman:2022ynf}
\begin{eqnarray}
{\rm Br} (B^0_s \to D^-_s \, \pi^+ \,)\Big|_{\rm exp.} & = & (2.98 \pm 0.14) \times 10^{-3}\,,
\label{eq:Bs-to-Ds-pi-Exp}
\\[1mm] 
{\rm Br} (B^0 \to D^- K^+)\Big|_{\rm exp.} & = & (2.05 \pm 0.08) \times 10^{-4}\,,
\label{eq:B-to-D-K-Exp}
\end{eqnarray}
based on measurements by the LHCb, Belle and CDF collaborations 
\cite{Belle:2001ccu, CDF:2006hob, Belle:2008ezn, LHCb:2013vfg, Belle:2021udv, LHCb:2021qbv}. 

On the theoretical side, the amplitude for the decays  
 $\bar B_{(s)}^0 \to D_{(s)}^+ L^-$, 
with $L = \{\pi, K\}$, can be computed by introducing the effective Hamiltonian
${\cal H}_{\rm eff}$, governing the tree-level non-leptonic $b$-quark transition $b \to~c \bar u q$, with $q =\{d, s \}$. This reads~\cite{Buchalla:1995vs}
\begin{equation}
	{\cal H}_{\rm eff} = \frac{G_F}{\sqrt 2} V_{cb} V_{uq}^*
	\Bigl(C_1 O_1^q + C_2 O_2^q \Bigr) + {\rm h. c.}\,,
\label{eq:Heff}
\end{equation}
where $G_F$ is the Fermi constant, $V_{q_1q_2}$ denote the Cabibbo-Kobayashi-Maskawa (CKM) matrix elements, and the $\Delta B = 1$ current-current operators $O_{1,2}^{q}$ are defined in the Chetyrkin-Misiak-M\"unz (CMM) basis respectively as \cite{Chetyrkin:1996vx, Chetyrkin:1997gb}\footnote{Note that the notation in eq.~\eqref{eq:O1-O2-def} is opposite to the one used in Refs.~\cite{Chetyrkin:1996vx, Chetyrkin:1997gb}.}
\begin{equation}
O_1^q  = 
\left(\bar c \Gamma^\mu b \right)
(\bar q \Gamma_\mu u)\,, 
\qquad 
O_2^q  =  
\left(\bar c \Gamma^\mu t^a  b \right)
(\bar q \Gamma_\mu t^a u),
\label{eq:O1-O2-def}
\end{equation}
with $\Gamma_\mu = \gamma_\mu (1 - \gamma_5)$,
and $t^a$ being the SU(3)$_c$ generators in the fundamental representation. 
In eq.~\eqref{eq:Heff}, the Wilson coefficients $C_{1,2}$ are evaluated at the renormalisation scale $\mu_b \sim m_b$, and are currently known up to NNLO logarithmic accuracy~\cite{Gorbahn:2004my}.
The amplitude then takes the form 
\begin{equation}
{\cal A} (\bar B_{(s)}^0 \to D_{(s)}^+ L^-) 
=
- \frac{G_F}{\sqrt 2} V_{cb} V_{uq}^* 
\Bigl(C_1 \langle O_1^q \rangle 
+ C_2 \langle O_2^q \rangle \Bigr)\,,
\label{eq:decay-ampl-2}
\end{equation}
where we have introduced the shorthand notation 
$\langle O_i^q \rangle \equiv \langle D^+_{(s)} L^-| O_i^q | \bar B_{(s)}^0 \rangle$.\ 

The simplest approach to determine the two matrix elements appearing in eq.~\eqref{eq:decay-ampl-2} is naive QCD factorisation~(NQCDF). Within this approximation, the matrix element of the colour-singlet operator factorises into the product of the light meson decay constant $f_L$ and of the scalar form factor $f_0^{B_{(s)}D_{(s)}}(q^2)$, parameterising the $B_{(s)} \to D_{(s)}$ transition, whereas the matrix element of the colour-octet operator vanishes, namely
\begin{align}
\langle O_{1}^q \rangle\Big|_{\rm NQCDF}=  
i f_{L} (m_{B_{(s)}}^2 - m_{D_{(s)}}^2) f_0^{B_{(s)}D_{(s)}} (m_L^2)\,,
\qquad \,
{\rm and}
\qquad \,
\langle O_2^q \rangle\Big|_{\rm NQCDF}= 0\,.
\label{eq:ME-NQCDF}
\end{align}
Because of eq.~\eqref{eq:ME-NQCDF}, in the literature, $\MEs$ and $\MEt$ are commonly referred to as the factorisable and non-factorisable matrix elements. Also in the present work we follow this notation, however with the remark that the distinction applies strictly only to LO-QCD, since, by including perturbative gluon corrections, both the matrix elements receive factorisable and non-factorisable contributions. We stress in fact that the accuracy of our study limits at LO-QCD and, unless explicitly stated, $\MEs$, $\MEt$ should always be intended as the corresponding tree-level matrix elements.

A first estimate of the non-factorisable matrix element $\MEt$ beyond the NQCDF approximation was obtained by Blok and Shifman in 1992~\cite{Blok:1992na}. Using the framework of QCD sum rule (QCDSR)~\cite{Shifman:1978bx, Shifman:1978by} with a two-point correlation function, the authors found positive non-factorisbale corrections of the order of few percent, to the amplitude in eq.~\eqref{eq:decay-ampl-2}. Specifically, with the NLO values $C_1 = 1.01$ and $C_2 =- 0.32$, their result leads to \footnote{Using the LO values for the Wilson coefficients $C_1= 1.03$ and $C_2 = - 0.53$, which corresponds to the accuracy of Ref.~\cite{Blok:1992na}, yields instead 
 $C_2 \langle O_2^d \rangle / C_1  \langle O_1^d \rangle  \sim 13 \%$. } 
\begin{equation}
\frac{C_2 \langle O_2^d \rangle}{C_1  \langle O_1^d\rangle}\Bigg|_{\rm QCDSR} \!\! \!\! \!\! \sim 8 \% \,, \qquad \qquad \bar B_s^0 \to D_s^+ \pi^-\,.
\label{eq:Blok-Shifman-result}
\end{equation}
It is worthwhile pointing out that a later study, of the theoretically less clean mode $\bar B^0 \to D^0 \pi^0$, was performed in Ref.~\cite{Halperin:1994hg}, using the light-cone sum rule (LCSR) method \cite{Balitsky:1989ry} with pion light-cone distribution amplitudes~(LCDAs). Also in the latter work, estimates of $\MEt$ gave a sizeable and positive result, in consistency with Ref.~\cite{Blok:1992na}.
However, a more recent analysis of the same decay $\bar B^0 \to D^0 \pi^0$ performed in Ref.~\cite{Cui:2004jc}, again with the LCSR framework and pion LCDAs, but starting from a three-point correlation function, closely following the approach introduced in Ref.~\cite{Khodjamirian:2000mi}, found the non-factorisable contribution to be sizeable, but negative. 

At the end of the '90s, a new framework for the computation of several non-leptonic two-body $B$-meson decays was developed in Refs.~\cite{Beneke:1999br, Beneke:2000ry, Beneke:2001ev}, the QCD factorisation (QCDF) method. Within QCDF, the matrix elements $\langle O_i^q \rangle$ in eq.~\eqref{eq:decay-ampl-2} can be computed respectively as
\begin{equation}
\langle O_{i}^q \rangle \Big|_{\rm QCDF} =    
\sum_{j} f_{j}^{B_{(s)} D_{(s)}} (m_L^2) \int\limits_0^1 \! d u \, T_{ij} (u) \varphi_L (u) + {\cal O} \left(\frac{\Lambda_{\rm QCD}}{m_b}\right),
\label{eq:QCDF-scheme}
\end{equation}
where $T_{ij} (u)$ are the corresponding hard-scattering kernels, which can be calculated perturbatively in QCD, $\phi_L (u)$ denotes the $L$-meson LCDA, and $f_{j}^{B_{(s)} D_{(s)}} (q^2)$ are the form factors parametrising the $B_{(s)} \to D_{(s)}$ transition. 
The latter two inputs are related to the hadronic structure of the mesons considered and therefore must be determined using some non-perturbative technique like Lattice QCD or QCD sum rule. In some cases, they could also be extracted from data.
It is important to emphasize that since the factorisation formula in eq.~\eqref{eq:QCDF-scheme} 
holds up to power corrections of the order of $\Lambda_{\rm QCD}/m_b$, the QCDF framework
allows one to systematically compute only the leading power contribution to the amplitude, however, to higher order in $\alpha_s$. Furthermore, the matrix element $\MEt$, vanishing at LO-QCD, constitutes at this order a purely next-to-leading power effect i.e.\ $\MEt = {\cal O}(\Lambda_{\rm QCD}/m_b) + {\cal O}(\alpha_s)$. 

The QCDF method was proven to be a very powerful tool for the computation of several non-leptonic $B$-meson decays. Remarkably, the hard-scattering kernels $T_{ij} (u)$ are known up to NNLO-QCD corrections~\cite{Huber:2016xod}. 
However, a systematic study of these processes beyond the leading power becomes challenging.
A recent analysis of the decays $\bar B_{(s)}^0 \to D_{(s)}^{(*)+} L^-$ within QCDF was performed in Ref.~\cite{Bordone:2020gao}. The authors have included NNLO-QCD corrections for the hard-scattering kernels from Ref.~\cite{Huber:2016xod}, and the $B_{(s)} \to D_{(s)}^{(*)}$ form factors from Ref.~\cite{Bordone:2019guc}, where the latter were obtained fitting the corresponding Isgur-Wise functions 
up to corrections of the order ${\cal O}\left(\Lambda_{\rm QCD}^2/m_c^2\right)$ in the Heavy Quark Expansion, by combining both Lattice QCD data \cite{FermilabLattice:2014ysv, MILC:2015uhg, Na:2015kha, Harrison:2017fmw, McLean:2019qcx, McLean:2019sds} and QCDSR results \cite{Gubernari:2018wyi, Bordone:2019guc}. 
In addition, they have also obtained a first estimate of the next-to-leading power corrections, by computing, within LCSR, the corresponding hadronic matrix element emerging in QCDF~\cite{Beneke:2000ry}.
This effect was found to be very small, of the order of 
sub-percent, namely~\cite{Bordone:2020gao}
\begin{equation}
\frac{{\cal A} (\bar B_{(s)}^0 \to D_{(s)}^+ L^-)|_{\rm NLP}}{{\cal A} (\bar B_{(s)}^0 \to D_{(s)}^+ L^-)|_{\rm LP}} \simeq - [0.06, 0.6] \% \,,
\label{eq:NLP-LP-QCDF}
\end{equation}
leading all together to very precise predictions for the 
branching fractions, which resulted to be significantly above the corresponding experimental data. Specifically, the authors of Ref.~\cite{Bordone:2020gao} have obtained
\begin{align}
{\rm Br} (\bar B^0_s \to D^+_s \, \pi^- \,)\Big|_{\rm QCDF} &=  (4.42 \pm 0.21) \times 10^{-3}\,, 
\label{eq:Bs-to-Ds-pi-QCDF}
\\[1mm]
{\rm Br} (\bar B^0 \to D^+ K^-)\Big|_{\rm QCDF}  & =  (3.26 \pm 0.15) \times 10^{-4}\,,
\label{eq:B-to-D-K-QCDF}
\end{align}
in clear tension with the values shown in eqs.~\eqref{eq:Bs-to-Ds-pi-Exp}, \eqref{eq:B-to-D-K-Exp} \footnote{Note that in the SM, the direct CP-asymmetry in these decays is negligible, therefore ${\rm Br} (\bar B_{(s)}^0 \to D_{(s)}^+ L^-) = {\rm Br} (B_{(s)}^0 \to D_{(s)}^- L^+) $. However, this might not necessarily hold in the presence of NP. In this respect, a clear experimental test was suggested in Refs.~\cite{Fleischer:2016dqd, Gershon:2021pnc}.}. Finally, a later study of the same decays within QCDF, however only at leading power, was performed in Ref.~\cite{Cai:2021mlt}. The conclusions obtained were similar to those in Ref.~\cite{Bordone:2020gao} and also their analysis revealed a large discrepancy with the data. 
This puzzling pattern has attracted significant attention in the recent literature, and has led to further investigations of these decays, both within the SM and beyond \cite{Beneke:2021jhp, Endo:2021ifc, Iguro:2020ndk, Cai:2021mlt, Bordone:2021cca, Fleischer:2021cct, Fleischer:2021cwb, Lenz:2022pgw, Gershon:2021pnc}. A conclusive explanation is, however, still missing.

The current status of the non-leptonic decays $\bar B_{(s)} \to D^+_{(s)} L^-$ represents a strong motivation to revisit the estimates of the non-factorisable contribution due to $\MEt$.  
Given the two very different results shown in eqs.~\eqref{eq:Blok-Shifman-result}, \eqref{eq:NLP-LP-QCDF},  
we present a new determination of the matrix element $\MEt$ within LCSR, starting from a three-point correlation function with $B$-meson LCDAs, partially following the method suggested in Ref.~\cite{Khodjamirian:2000mi}. Moreover, we also compute for the first time within LCSR, the factorisable matrix element~$\MEs$, including both two- and three-particle LCDAs, and thus obtain predictions for the corresponding branching fractions entirely within the same framework. 

The paper is organised as follows. In section~\ref{sec:ME-Ot-LCSR}, we describe the computation of the non-factorisable matrix element $\MEt$ within LCSR. More precisely, a detailed derivation of the operator product expansion for the three-point correlator is presented in section~\ref{sec:Der-OPE}, the light-cone dominance of the correlation function and the problem associated with the lack of
generalised $B$-meson quark-gluon-quark matrix elements with non-aligned fields, are discussed in section~\ref{sec:LC-dominance}, while the hadronic dispersion relations are derived in section~\ref{sec:Dispersion-relations}.    
In section~\ref{sec:ME-O1-QCDSR}, we briefly discuss the computation of the factorisable matrix element $\MEs$ within LCSR. Our numerical analysis is presented in section~\ref{sec:Num-analysis}. In particular, a detailed discussion of the inputs used in the analysis can be found in section \ref{sec:Inputs}, while our results are shown in section~\ref{sec:Results}. Finally, in section~\ref{sec:conclusion}, we present our conclusions, as well as a comprehensive outlook for future improvements.

\section{\boldmath Determination of $\langle O_2 \rangle$ from LCSR}
\label{sec:ME-Ot-LCSR}
\subsection{Derivation of the OPE for the correlator}
\label{sec:Der-OPE}
To compute the hadronic matrix element $\langle O_2 \rangle$ \footnote{Unless explicitly stated, we assume, for definiteness, the mode $\bar B_s \to D_s^+ \pi^-$, and often drop, for the sake of a cleaner notation, all labels. The discussion presented here, in fact, straightforwardly extends to the mode $\bar B^0 \to D^+ K^-$, once the proper replacements are taken into account.} within the framework of LCSR, we start by introducing the following three-point correlation function
\begin{equation}
F^{\, O_2}_\mu (p, q) = i^2 \int d^4 x \, e^{i p \cdot x} \int d^4 y
\, e^{i q \cdot y} \,
\langle 0| {\rm T} \! \left\{j^{D}_5 (x), 
O_2 (0), j^\pi_\mu (y) \right\} 
| \bar B (p + q) \rangle\,,
\label{eq:Correlator-O1T}
\end{equation}
where $j^{D}_5 (x) = i m_c \, \bar s \gamma_5 c$
and $j^\pi_\mu (y) = \bar u \gamma_\mu \gamma_5 d $
are suitable interpolating currents with the quantum numbers of the $D^+_{s}$- and $\pi^-$-mesons, and momenta $p^\mu$ and~$q^\mu$, respectively. 
We consider eq.~\eqref{eq:Correlator-O1T} in the kinematical domain $P^2 \equiv -p^2 \gg \Lambda^2,$ and $Q^2 \equiv -q^2 \gg \Lambda^2$, with $\Lambda$ denoting a small hadronic scale of the order of $\Lambda_{\rm QCD}$. With this choice, as discussed further in section~\ref{sec:LC-dominance}, the dominant contribution to the correlator originates from the region in which both $x^\mu$ and $y^\mu$ are approximately light-like and aligned along different light-cone (LC) directions, i.e.\
\begin{equation}
x^2 \sim 0\,, \qquad y^2 \sim 0\,, \qquad (x-y)^2 \not \sim 0\,.
\end{equation}
A double LC expansion, however, currently can not be consistently performed due to the lack of the proper hadronic input functions, that is of the $B$-meson three-particle non-local matrix element with the gluon and the spectator quark aligned on different LC directions.
For this reason, in the following, we consider the specific case of 
LC-local dominance, which is also compatible with the present kinematics, see section~\ref{sec:LC-dominance}, and expand the time-ordered product in eq.~\eqref{eq:Correlator-O1T} around $x^2\sim 0$ but $y^\mu \sim 0$ \footnote{In principle, also the opposite choice i.e.\ expanding around $y^2\sim 0 $ and $x^\mu \sim 0$ could be considered. We leave the investigation of this alternative scenario for a future study.}. In this way, in fact, the relevant hadronic matrix element can be derived from the expression for aligned fields given e.g.\ in Ref.~\cite{Braun:2017liq}, by setting the LC coordinate of the gluon field to zero. We return to this point later on.

Expanding the time-ordered product in eq.~\eqref{eq:Correlator-O1T}, we thus obtain
\begin{align}
F^{\, O_2}_\mu(p,q) & = - i m_c \int  d^4 x  \int d^4 y  \,\, e^{i p\cdot x} \, e^{i q \cdot y}  \,\, \langle 0| \bar s^i(x)\gamma_5 \, i S^{(c)}_{ij}(x,0) \gamma_\rho (1-\gamma_5)  
\nonumber \\[1mm]
& \times i S^{(u)}_{mn}(0,y) \gamma_\mu \gamma_5 \, i S_{nl}^{(d)}(y,0) \gamma^\rho (1-\gamma_5) b^k (0) |\bar B(p+q) \rangle \, t^a_{jk} t^a_{lm}\,,
\label{eq:Correlator-O1T-2}
\end{align}
where $S_{ij}(x,y)$ denotes the corresponding quark propagator, with the specific quark indicated in the superscript.
In deriving eq.~\eqref{eq:Correlator-O1T-2}, the operator $O_2$ has been Fierz-transformed to avoid the computation of traces involving $\gamma_5$ in dimensional regularisation. Note that this can be consistently done since, with the present choice of the operator basis, the Fierz symmetry is respected also at the one-loop order, see e.g.~Ref.~\cite{Buras:1989xd}.
\begin{figure}[t]
    \centering
    \includegraphics[scale=0.5]{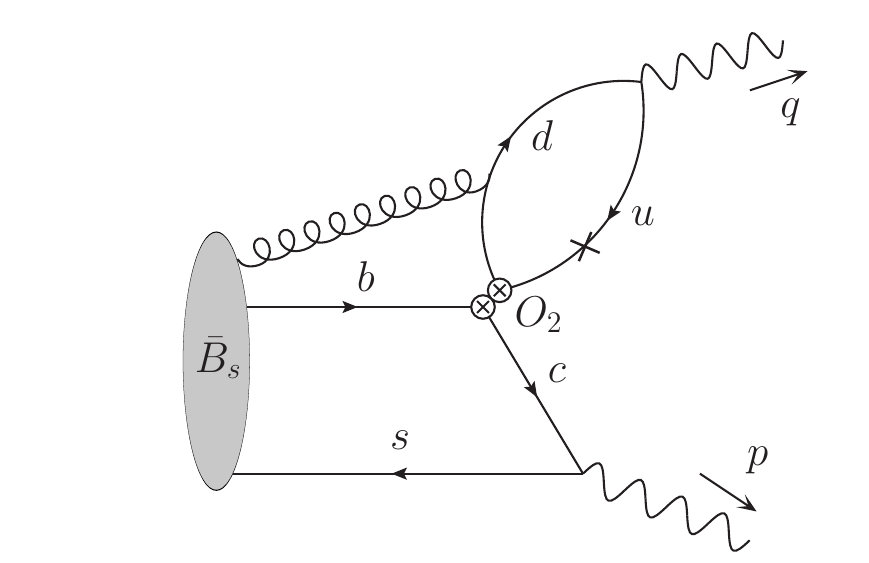}
    \caption{Diagram describing the leading contribution in the OPE for the correlator $F^{\, O_2}_\mu (p, q)$. The cross indicates the second possible point of gluon emission.}
    \label{fig:correlator-OPE-O2}
\end{figure}
Owing to the colour structure of eq.~\eqref{eq:Correlator-O1T-2}, the first non-vanishing contribution corresponds to the emission of one gluon from either the $u$- or $d$-quark propagators, as shown in Figure~\ref{fig:correlator-OPE-O2}. 
In the Fock-Schwinger gauge, the local expansion of the quark propagator in an external background gluon field, including the leading one-gluon corrections, can be found e.g.\ in Refs.~\cite{Novikov:1983gd, Balitsky:1987bk, Belyaev:1994zk}. The corresponding expression, in the case of massless quark, i.e.\ for $q=\{u,d\}$, takes the following form  
\begin{align}
    S^{(q)}_{ij}(x,y) = \! \int \! \frac{d^4 k}{(2 \pi)^4} e^{-i k (x-y)} \left[ \frac{\delta_{ij} \, \slashed k }{k^2  + i \varepsilon} - \frac{G^{a}_{\mu \nu} t^a_{ij}}{4}  \frac{(\slashed k \, \sigma^{\mu \nu} + \sigma^{\mu \nu} \slashed k)}{(k^2 + i\varepsilon)^2}  \right] + \ldots \,,
    \label{eq:light-quark-prop}
\end{align}
where $G_{\mu \nu}$ is the gluon field strength tensor evaluated at the origin, $\sigma_{\mu \nu} = (i/2)[\gamma_\mu, \gamma_\nu]$, and the ellipses denote terms of higher order neglected at the current accuracy. 
Substituting eq.~\eqref{eq:light-quark-prop} into eq.~\eqref{eq:Correlator-O1T-2}, the integral over $y^\mu$ can be easily calculated \footnote{Manipulations involving the Dirac algebra are performed using the Mathematica package $\mathtt{ FeynCalc}$~\cite{Shtabovenko:2020gxv}.}. This yields
 \begin{align}
     F^{\, O_2}_\mu (p,q) = - \frac{m_c}{2} \int d^4 x \, e^{i p \cdot x} \,\langle 0| \bar s^i (x) \gamma_5 S_0^{(c)} (x) G_{\nu \rho}^a (0) t^a_{ij} \, {\cal I}^{\,\, \nu \rho}_{\mu}(q) b^j (0)| \bar B (p+q)\rangle,
     \label{eq:F-tilde-1}
 \end{align}
 with $S_0^{(c)}(x)$ denoting the free charm-quark propagator, and the tensor ${\cal I}_{\mu \nu \rho}$ being
 \begin{align}
     {\cal I}_{\mu \nu \rho} (q) = \frac{i}{4 \pi^2 (q^2 + i \varepsilon)} \Bigl( q_\nu \, q^\lambda \epsilon_{\mu \rho \tau \lambda}-
     q_\mu \, q^\lambda \epsilon_{\nu \rho \tau \lambda} -
     q^2 \epsilon_{\mu \nu \rho \tau}\Bigr) 
     \gamma^\tau (1 - \gamma_5)\,.
     \label{eq:I-alphabetamu}
 \end{align}
The result in eq.~\eqref{eq:I-alphabetamu} has been obtained by computing the corresponding loop integral in naive dimensional regularisation (NDR) with $d = 4 - 2 \epsilon$ and anticommuting prescription for~$\gamma_5$. We note that the divergent $1/\epsilon$ contributions exactly cancel when considering the gluon emission from both the $u$- and $d$-quark propagators, leading to a finite expression, in consistency with Refs.~\cite{Khodjamirian:2010vf, Gubernari:2020eft}. 
In addition, we have also performed the  computation of the loop function in eq.~\eqref{eq:I-alphabetamu} using the explicit coordinate representation of the local expansion of the propagator, details of which are given in appendix~\ref{app:loop-func-coord-space}. 

To proceed with the calculation of $F^{\, O_2}_\mu$, we must evaluate the non-local matrix element appearing in eq.~\eqref{eq:F-tilde-1}.
At leading order in the heavy quark effective theory~(HQET), the non-local vacuum-to-$B$ three-particle matrix element with the gluon and the spectator quark aligned on the same light-cone direction can be parametrised in terms of eight LCDAs~\cite{Geyer:2005fb}. The matrix element in eq.~\eqref{eq:F-tilde-1} corresponds to the specific configuration in which the gluon field is fixed at the origin, and its parametrisation can be derived by taking the local limit of the result given e.g.\ in Ref.~\cite{Braun:2017liq}. 
We present below only the final expression and refer to appendix~\ref{app:B} for the intermediate steps. At leading order in HQET, we then have
\begin{align}
    & \langle 0| \bar s_\alpha (x) G_{\mu \nu}(0) b_{\beta}(0)|\bar B(p+q) \rangle = \frac 1 2 F_{B}(\mu) \sqrt{m_{B}}
    \int_0^\infty \!\! d \omega_1 \, e^{- i \omega_1 v \cdot x}
    \nonumber \\[1mm]
    &\qquad \times \Big\{P_+ \big[(v_\mu \gamma_\nu - v_\nu \gamma_\mu) (\hat \psi_A - \hat \psi_V)  - i \sigma_{\mu \nu} \hat \psi_V - i (x_\mu v_\nu - x_\nu v_\mu) \bar{\hat{\psi}}_{X_A} 
    \nonumber \\[1mm] 
    &
    \qquad \qquad
    + i (x_\mu \gamma_\nu - x_\nu \gamma_\mu) (\bar{\hat{\psi}}_W + \bar{\hat{\psi}}_{Y_A}) 
    - \epsilon_{\mu \nu \eta \tau} x^\eta v^\tau \gamma_5 
    \bar{\hat{\psi}}_{\tilde{X}_A}  + \epsilon_{\mu \nu \eta \tau} x^\eta \gamma^\tau \gamma_5  \, \bar{\hat{\psi}}_{\tilde{Y}_A}
    \nonumber \\[1mm]
    & \qquad \qquad
    + (x_\mu v_\nu - x_\nu v_\mu) \slashed x \, \bar{\bar{\hat{ \psi}}}_W 
    - (x_\mu \gamma_\nu - x_\nu \gamma_\mu) \slashed x \, \bar{\bar{\hat{\psi}}}_ Z
    \big] \gamma_5 \Big\}_{\beta \alpha}(\omega_1 ; \mu) \,,
    \label{eq:3-part-ME}
\end{align}
where $\alpha, \beta,$ are spinor indices, $v^\mu =( p^\mu + q^\mu)/m_{B}$ is the velocity of the $B$ meson, $F_{B}(\mu)$ is the HQET decay constant, and $P_+ = (1 + \slashed v)/2$. 
Three comments are in order with respect to Ref.~\cite{Braun:2017liq}. First, the terms proportional to $\epsilon_{\mu \nu \eta \tau}$ appear with an opposite sign because of the different convention adopted in our work for the Levi-Civita tensor, namely $\varepsilon^{0123} = +1$, see also appendix~\ref{app:A}. Second, we have relabelled some LCDAs to make the notation throughout the paper more transparent. Third, the extra mass factor in eq.~\eqref{eq:3-part-ME} follows from the conversion from HQET to QCD for the $B$-meson state.
Moreover, we have also introduced the notation \footnote{The $\mu$-dependence of the LCDAs is often omitted, however it should always be understood.}
\begin{equation}
    \bar{\hat{\psi}}(\omega_1) = 
    \int_0^{\omega_1} \!\! d \eta \, 
      \hat\psi(\eta) \,,
\qquad
    \bar{\bar{\hat{\psi}}}(\omega_1) = 
    \int_0^{\omega_1} \!\! d \eta
    \int_0^{\eta} \!\! d \eta^\prime \, 
      \hat\psi(\eta^\prime) \,.
\end{equation}
Given the explicit $x$-dependence of eq.~\eqref{eq:3-part-ME}, the integration over $x^\mu$ in eq.~\eqref{eq:F-tilde-1} can be now performed. To this end, it appears to be more convenient to use the coordinate representation of the free charm-quark propagator, which reads
\begin{equation}
	S^{(c)}_{0} (x) =  
	- \frac{i m_c^2}{4 \pi^2} \left[ \frac{K_1 (m_c \sqrt{-x^2})}{\sqrt{-x^2}} -
	i \frac{\slashed x}{x^2} K_2(m_c \sqrt{-x^2}) \right],  
 \label{eq:prop-coordinate-space}
\end{equation}
with $K_n (z)$ being the modified Bessel function of the second kind of order $n$. Taking into account eqs.~\eqref{eq:3-part-ME}, \eqref{eq:prop-coordinate-space}, we are then left with the evaluation of tensor integrals of the type
\begin{equation}
    \int d^4 x \, e^{i \tilde p \cdot x}  \, \frac{K_1(m_c \sqrt{-x^2})}{\sqrt{- x^2}}\,  \Big\{ 1, x^\mu, x^\mu x^\nu, \ldots  \Big\} \,, 
    \label{eq:int-x-K1}
\end{equation}
\begin{equation}
    \int d^4 x \,  e^{i \tilde p \cdot x}  \, \frac{K_2(m_c \sqrt{-x^2})}{x^2}\,  \Big\{x^\mu, x^\mu x^\nu, x^\mu x^\nu x^\rho, \ldots \Big\} \,, 
    \label{eq:int-x-K2}
\end{equation}
 where, for simplicity, we have introduced the compact notation $\tilde p^\mu = p^\mu - \omega_1 v^\mu $. The result for the inverse Fourier transform of Bessel functions in eqs.~\eqref{eq:int-x-K1}, \eqref{eq:int-x-K2}, is explicitly given in appendix~\ref{app:D}. Using eqs.~\eqref{eq:int-K1-app}-\eqref{eq:int-K2-x3-app}, we then arrive at the final form of the correlator in eq.~\eqref{eq:Correlator-O1T}, that is
\begin{equation}
    F^{\, O_2}_\mu(p,q) = \left((p \cdot q) \, q_\mu  - q^2 p_\mu \right) F^{\, O_2} (p^2, q^2)\,,
    \label{eq:F-tilde-transverse}
\end{equation}
with $F^{\, O_2} (p^2, q^2)$ denoting the corresponding Lorentz invariant amplitude. On this point, an important remark is in order. The result for the correlation function in eq.~\eqref{eq:F-tilde-transverse} is transversal with respect to the momentum of the light-quark current $q^\mu$, as expected, since in the limit of massless $u$- and $d$-quark, the axial-vector current~$j_\mu^\pi$ must be conserved~\footnote{Since we neglect the mass of the strange quark in the loop, the same argument applies also to the decay $\bar B^0 \to D^+ K^-$.}. However, when trying to compute the correlator in eq.~\eqref{eq:Correlator-O1T} by expanding the time-ordered product around $x^2 \sim 0$, $y^2 \sim 0$, and by using the expression for the $B$-meson three-particle matrix element with both the gluon and the spectator quark aligned on the same light-cone direction, i.e.\ implicitly assuming that also $(x-y)^2 \sim 0$, we obtain an expression for $F^{\, O_2}_\mu$ which is not transversal \footnote{Specifically, we find that the transversality of the correlator is violated by terms proportional to $u\,  \omega_2/m_B$, with $\omega_2$ being the momentum of the gluon field and $u \in [0,1]$ a LC parameter. We have also explicitly checked that these terms do not vanish in the final result, i.e.\ after integration over $u$ and~$\omega_2$.}. In this respect, we also note that  in the case of charm loop with photon coupling studied e.g.\ in Refs.~\cite{Khodjamirian:2010vf, Gubernari:2020eft}, the expression of the non-local amplitude due to soft gluon emission appears actually to be not transversal with respect to the photon momentum.
Surprisingly, this has not been pointed out in the above references, nor, to our best knowledge, elsewhere in the literature. 
Further investigations of this issue would clearly be of utmost importance not only to improve the current estimate of the non-factorisable amplitude in non-leptonic $B$-meson decays, but also in light of the impact that a better understanding of these non-local effects could have on the present status of the $B$ anomalies, see e.g.\ the reviews~\cite{Albrecht:2021tul, London:2021lfn}. 

Returning to eq.~\eqref{eq:F-tilde-transverse}, we isolate the coefficients of the two Lorentz structures and rewrite 
\begin{equation}
F^{\, O_2}_\mu(p,q) =  
F^{\, O_2}_q (p^2,q^2) \, q_\mu +
F^{\, O_2}_p (p^2,q^2) \, p_\mu,
\label{eq:Fq-Fp}
\end{equation}
where the LC-local operator product expansion (OPE) for the amplitude $F^{\,O_2}_q (p^2,q^2)$, relevant for the hadronic dispersion relations, see section~\ref{sec:Dispersion-relations}, can be expressed as
\begin{equation}
    \big[F^{\, O_2}_{q}(p^2,q^2)\big]_{\rm OPE}= F_B \sqrt{m_B} \, m_c \int \limits_0^\infty d \omega_1 \sum_{\hat \psi} \hat \psi(\omega_1) \sum_{n= 1}^{3}  \frac{c_n^{\hat \psi}(\omega_1, q^2)}{(q^2+ i \varepsilon) \big[ \tilde s(\omega_1, q^2) - p^2 - i \varepsilon \big]^n}  \,.
    \label{eq:Fq-OPE}
\end{equation}
In the above equation $\hat \psi = \hat \psi_A, \hat \psi_V,\ldots,$ and for later convenience, the coefficients of the LCDAs have been suitably manipulated so that in eq.~\eqref{eq:Fq-OPE} the dependence on~$p^2$ is contained exclusively in the denominators. Finally, the function $\tilde s(\omega_1, q^2)$ reads 
\begin{equation}
    \tilde s(\omega_1, q^2) = \left( \frac{m_{B}}{m_{B}- \omega_1} \right) \Big[ m_c^2 + \omega_1 m_{B} - q^2 \frac{\omega_1}{m_{B}}- \omega_1^2 \Big]\,,
    \label{eq:tilde-s-func}
\end{equation}
while the analytic expressions of the OPE coefficients $c_n^{\hat \psi}(\omega_1, q^2)$ can be found in appendix~\ref{app:E}.
\subsection{Light-cone dominance of the correlator}
\label{sec:LC-dominance}
In this section we investigate the conditions for the LC dominance of the correlation function in eq.~\eqref{eq:Correlator-O1T} and discuss the corresponding kinematics. The correlator $F_\mu^{\, O_2}$, in fact, describes the decay of a heavy $B$ meson into two currents with momenta $p^\mu$ and $q^\mu$, namely
\begin{equation}
m_{B} v^\mu = p^\mu + q^\mu\,,
\label{eq:decay-Bmeson}
\end{equation}
where $v^\mu = p^\mu_{B}/m_{B}$ is the $B$-meson velocity. In order to be far away from hadronic thresholds originating from the two interpolating currents, we consider the kinematical region in which
\begin{equation}
Q^2 \sim P^2 \sim m_{B}  \Lambda,  \qquad 
P^2 \equiv - p^2\,, \qquad Q^2 \equiv - q^2\,,
\label{eq:Psq-Qsq-scaling}
\end{equation}
with $\Lambda$ being a small non-perturbative scale of the order of $\Lambda_{\rm QCD}$. Hence, both $p^2$ and $q^2$ are assumed to be space-like and large, leading to the following power counting
\begin{equation}
m_{B}^2 \gg Q^2 \sim P^2 \gg \Lambda^2\,.
\end{equation}
It is convenient to study eq.~\eqref{eq:decay-Bmeson} in the rest frame of the $B$ meson, i.e.\ $v^\mu = (1, \vec 0 \,)$, aligning, for simplicity, the $z$-axis along the direction of the decay. Furthermore, we introduce the two light-cone vectors $n_{\pm}^\mu$, with $n_+^2 =  n_-^2 = 0$, such that $v^\mu = (n_+^\mu + n_-^\mu)/2$. Specifically
\begin{equation}
    n_+^\mu = (1, 0, 0 , 1)\,, \qquad n_-^\mu = (1, 0, 0, -1)\,, \qquad (n_+ \cdot n_-) = 2\,.
\end{equation}
A solution for $p^\mu$ and $q^\mu$, up to corrections of the order $P^4/m_B^4$ and $Q^4/m_B^4$, is given~by~\footnote{Eq.~\eqref{eq:decay-Bmeson} admits also a second solution obtained by exchanging the coefficients of $n_-^\mu$ and $n_+^\mu$.
Without loss of generality, however, we parametrise the momenta according to eq.~\eqref{eq:momenta-solution}.}
\begin{equation}
\left\{ 
\begin{matrix}
\displaystyle{p^\mu = \left( \frac{m_B^2 + Q^2}{2m_B}\right) n_+^\mu +  \left(-\frac{P^2}{2 m_B}\right)   n_-^\mu} \,, 
\\[5mm]
\displaystyle{q^\mu = \left(-\frac{Q^2}{2 m_B}\right)  n_+^\mu +  \left( \frac{m_B^2 + P^2}{2m_B}\right)   n_-^\mu} \,, 
\label{eq:momenta-solution}
\end{matrix}
\right.
\end{equation}
where, due to our choice of the coordinate system, the components transversal to the light-cone vectors vanish, namely $p^\mu_\perp = q^\mu_\perp =0$. From eqs.~\eqref{eq:Psq-Qsq-scaling}, \eqref{eq:momenta-solution}, it then follows that whereas $p^\mu$ has a large component along $n_+^\mu$ and a small component along $n_-^\mu$, since the two coefficients respectively scale as $(p \cdot n_-) \sim m_B$, $(p \cdot n_+) \sim - \Lambda$, the behaviour is opposite for the two components of $q^\mu$, i.e.\ $(q \cdot n_-) \sim - \Lambda$, $(q \cdot n_+) \sim m_B$.

Having fixed the kinematics, we can turn to discuss the structure of the correlation function~$F_\mu^{O_2}$. The integrals in eq.~\eqref{eq:Correlator-O1T} are dominated by the values of $x^\mu$ and~$y^\mu$ in correspondence of which the argument of the exponentials is not large \footnote{This follows from the Riemann-Lebesgue theorem.}.
With the choice of momenta in eq.~\eqref{eq:momenta-solution}, the absence of fast oscillations, see also e.g.\ Refs.~\cite{Colangelo:2000dp, Khodjamirian:2020btr} for details, is ensured given that
\begin{equation}
\left\{ 
\begin{array}{l}
\displaystyle{
\exp\{i p \cdot x\} \simeq \exp\{i \underbrace{m_B \, x_0/2}_{\lesssim {\cal O}(1)} \} \, \exp\{- i \underbrace{\left(m_B + 2 \Lambda \right) x_3/2}_{\lesssim {\cal O}(1)}\}}\,,
\\[8mm]
\displaystyle{
\exp\{i q \cdot y\} \simeq \exp\{i \underbrace{m_B \, y_0/2}_{\lesssim {\cal O}(1)} \} \, \exp\{i \underbrace{\left(m_B + 2 \Lambda \right) y_3/2}_{\lesssim {\cal O}(1)}\}}\,,
\end{array}
\right.
\end{equation}
yielding respectively the bounds
\begin{equation}
\left\{ 
\begin{matrix}
\displaystyle
|x_0| \lesssim \frac{2}{m_B}\,, \quad |x_3| \lesssim \frac{2}{m_B + 2 \Lambda}\,,
\\[5mm]
\displaystyle
|y_0| \lesssim \frac{2}{m_B}\,, \quad |y_3| \lesssim \frac{2}{m_B + 2 \Lambda}\,.
    \end{matrix}
 \right.   
\label{eq:Local-dominance}
\end{equation}
From eq.~\eqref{eq:Local-dominance} it then follows that \footnote{The lower bound for $x^2$ and $y^2$ follows from the causality property of correlation functions, see e.g.\ Refs.~\cite{Iagolnitzer:1991wj, Muta:2010xua, Ruiz:2023ozv}.}
\begin{equation}
\left\{ 
\begin{matrix}
\displaystyle
x_0^2 - x^2_3 \lesssim \frac{4}{m_B^2}
\leq \frac{4}{m_B^2} + x_1^2 + x_2^2
\,, 
\\[5mm]
\displaystyle
y_0^2 - y^2_3 \lesssim \frac{4}{m_B^2}
\leq \frac{4}{m_B^2} + y_1^2 + y_2^2
\,, 
\end{matrix}
 \right. 
 \quad 
 \Rightarrow
 \quad
 \left\{ 
\begin{matrix}
\displaystyle
0 \leq x^2
\lesssim \frac{4}{m_B^2}
\,, 
\\[5mm]
\displaystyle
0 \leq y^2
\lesssim \frac{4}{m_B^2}
\,,
\end{matrix}
 \right. 
\end{equation}
showing that the region in which the time-ordered product in eq.~\eqref{eq:Correlator-O1T} dominates, corresponds to both $x^\mu$ and $y^\mu$ being approximately on the light-cone, i.e.\
\begin{equation}
x^2 \sim 0\,, \qquad y^2 \sim 0\,.
 \label{eq:LC-LC-dominance}
\end{equation}
On the other side, expressing the integrals in terms of light-cone coordinates, the exponentials in eq.~\eqref{eq:Correlator-O1T} read
\begin{equation}
\left\{ 
\begin{matrix}
\displaystyle
\exp\{i p \cdot x\} \simeq \exp\{-i \underbrace{\Lambda (x \cdot n_-)/2}_{\lesssim {\cal O}(1)} \} \,\exp\{i \underbrace{(m_B + \Lambda) (x \cdot n_+)/2}_{\lesssim {\cal O}(1)} \}\,,
\\[8mm]
\displaystyle
\exp\{i q \cdot y\} \simeq \exp\{ i \underbrace{(m_B + \Lambda) (y \cdot n_-)/2}_{\lesssim {\cal O}(1)} \} \, \exp\{ - i  \underbrace{\Lambda (y \cdot n_+)/2}_{\lesssim {\cal O}(1)} \}\,,
\end{matrix}
\right.
\end{equation}
and the absence of fast oscillations now leads to the conditions
\begin{equation}
\left\{ 
\begin{matrix}
\displaystyle
\frac{|x \cdot n_-|}{2} \lesssim \frac{1}{\Lambda}\,, \quad \frac{|x \cdot n_+|}{2} \lesssim \frac{1}{m_B + \Lambda}\,,
\\[5mm]
\displaystyle
\frac{|y \cdot n_+|}{2} \lesssim \frac{1}{\Lambda}\,,
\quad
\frac{|y \cdot n_-|}{2} \lesssim \frac{1}{m_B + \Lambda}\,.
    \end{matrix}
 \right.   
\label{eq:opposite-LC-dominance}
\end{equation}
Eq.~\eqref{eq:opposite-LC-dominance} thus shows that whereas the $x$-component along $n^\mu_-$ is strongly suppressed, the behaviour is opposite for $y^\mu$, meaning that the integrals in eq.~\eqref{eq:Correlator-O1T} are actually dominated by the region in which $x^\mu$ and $y^\mu$ are approximately aligned along different light-cone directions, namely\footnote{From eqs.~\eqref{eq:LC-LC-dominance}, \eqref{eq:opposite-LC-dominance}, it also follows that $|x \cdot n_+| \ll |x_\perp| \ll |x \cdot n_-|$ and $|y \cdot n_-| \ll |y_\perp| \ll |y \cdot n_+|$, where we have introduced the notation $a_\perp^\mu \equiv  a_\perp n_\perp^\mu$ with $n_\perp^2 = - 1$.
}
\begin{equation}
x^\mu \sim \frac{(x \cdot n_-)}{2} n^\mu_+ \,, \qquad y^\mu \sim \frac{(y \cdot n_+)}{2} n^\mu_- \,, \qquad (x-y)^2 \not \sim 0\,.
\label{eq:scaling-x-y}
\end{equation} 
Had we used the light-cone expansion of the propagator, instead of its local limit given in eq.~\eqref{eq:light-quark-prop}, 
the resulting matrix element would be $\langle 0 | \bar s_\alpha(x) G_{\mu \nu}(uy) b_\beta(0)| \bar B(p+q) \rangle$.
Due to eq.~\eqref{eq:scaling-x-y}, the computation of the time-ordered product in eq.~\eqref{eq:Correlator-O1T} in terms of a double LC expansion would thus require the knowledge of the $B$-meson quark-gluon-quark matrix element with non-aligned fields, which, as already stressed, is not yet available in the literature for generic Dirac structures~\footnote{Non-local $B$-meson matrix elements with non-aligned fields have been investigated in e.g.\ Refs.~\cite{Benzke:2010js, Bell:2013tfa, Qin:2022rlk}. In particular, vacuum-to-$B$ three-particle matrix elements with the gluon and the light spectator quark aligned on different light-cone directions have been discussed in Ref.~\cite{Qin:2022rlk}. In the latter reference, the authors have also proposed a parameterisation for the novel soft function corresponding to the matrix element $\langle 0 | \bar q (z_1 n_+) G_{\mu \nu} (z_2 n_-) n_-^\nu \slashed n_+ \gamma_\perp^\mu \gamma_5 h_v(0)| \bar B \rangle$.}. 
In this connection, we note that by using the $B$-meson three-particle matrix element with aligned fields, as previously done in similar computations, see e.g.\ Refs.~\cite{Khodjamirian:2010vf, Gubernari:2020eft}, 
one might miss the actual dominant contributions and obtain potentially incomplete results. 
This issue was also recently pointed out in Refs.~\cite{Melikhov:2022wct, Qin:2022rlk, Melikhov:2023pet}. 
Hence, since the local limit $y^\mu \sim 0$ is also compatible with the present kinematics, as it follows from eq.~\eqref{eq:Local-dominance} \footnote{We note that 
e.g.\ in the first study by Blok and Shifman of the non-leptonic decays here considered~\cite{Blok:1992na}, or in the determinations of the pion decay constant from QCDSR~\cite{Shifman:1978bx, Shifman:1978by}, the local expansion of the light-quark propagator is used in correspondence of a typical scale of $Q^2 \sim 1 \GeV^2$, which is consistent with our kinematics, cf.~eq.~\eqref{eq:Psq-Qsq-scaling}.}
, we have chosen to perform instead a LC-local expansion, which, albeit less accurate than a double LC expansion, allows us to circumvent the problem associated with the lack of the corresponding matrix element and to compute the correlation function in terms of known hadronic input functions without incurring potential inconsistencies.
\subsection{Hadronic dispersion relations and sum rule}
\label{sec:Dispersion-relations}
The OPE result in eq.~\eqref{eq:Fq-OPE} must now be linked to $\langle O_2 \rangle$, the matrix element we aim to estimate. To this end, we proceed with the derivation of the hadronic dispersion relations for the correlator $F_\mu^{\, O_2}$.
Starting with the $p^2$-channel, we insert into eq.~\eqref{eq:Correlator-O1T} a complete set of intermediate states with the $D_s^+$-meson quantum numbers. This gives
\begin{equation}
F_\mu^{\, O_2}(p, q) =  \frac{m_{D}^2 f_{D}}{m_{D}^2 - p^2} \hat F_\mu^{\, O_2}(p_D, q) + q_\mu \! \int\limits_{s_h^{(D)}}^\infty  d s \,\, \frac{\rho_h (s, q^2)}{s - p^2} + \ldots \,,
\label{eq:Disp-rel-psq}
\end{equation}
where the decay constant of the $D_s$ meson is defined as $\langle 0 |j_5^D| D \rangle= m_{D}^2 f_{D}$, and we have introduced the two-point correlation function $\hat F_\mu^{\, O_2}(p_D, q)$, describing the transition of a $\bar B_s$-meson into a $D_s^+$-meson and a current $j_\mu^\pi$, namely
\begin{equation}
\hat F_\mu^{\, O_2}(p_D, q) = i \int  d^4 y \,\, e^{i q \cdot y}\, \langle D(p_D)| {\rm T}\{O_2(0), j_\mu^\pi(y)\} |\bar B(p_D+q)\rangle\,,
\label{eq:two-point-correl}
\end{equation}
with $p_D^2 = m_D^2$.
In eq.~\eqref{eq:Disp-rel-psq}, the spectral density $\rho_h(s,q^2)$ accounts for the contribution of excited states and of the continuum in the $p^2$-channel, with $s_h^{(D)}$ indicating the lowest hadronic threshold. Note that, taking into account the Lorentz decomposition shown in eq.~\eqref{eq:Fq-Fp}, we have already isolated the coefficient
of $q_\mu$, which is the only one relevant for the final result, and that the ellipses in eq.~\eqref{eq:Disp-rel-psq} denote the remaining contribution proportional to $p_\mu$. As the complicated structure of the spectral density is in general difficult to determine, the integral on the r.h.s.\ of eq.~\eqref{eq:Disp-rel-psq} is often estimated by recurring to the principle of quark-hadron duality (QHD), see e.g.\ Ref.~\cite{Shifman:2000jv}. 
By analytically continuing the function $\big[F^{\, O_2}_{q}(p^2,q^2)\big]_{\rm OPE}$ in eq.~\eqref{eq:Fq-OPE} in the complex $p^2$-plane, we can express it in the form of a dispersive integral as
\begin{equation}
\big[F^{\, O_2}_{q}(p^2,q^2)\big]_{\rm OPE} =  \frac{1}{\pi} \int \limits_{m_c^2}^\infty ds \, \, \frac{{\rm Im}_s\big[F^{\, O_2}_{q}(s,q^2)\big]_{\rm OPE}}{s - p^2}\,,
\label{eq:ImFope-integral}
\end{equation}
with $m_c^2$ being the fist pole on the real axis $p^2 = s$. Using QHD, we thus approximate 
\begin{equation}
   \int\limits_{s_h^{(D)}}^\infty d s \,\, \frac{\rho_h (s, q^2)}{s - p^2} =  \frac{1}{\pi} \int \limits_{s_0^D}^\infty ds \, \, \frac{{\rm Im}_s\big[F^{\, O_2}_{q}(s,q^2)\big]_{\rm OPE}}{s - p^2}\,,
   \label{eq:QHD}
\end{equation}
valid at sufficiently large and negative values of $p^2$. Here, $s_0^D$ is an effective threshold parameter to be determined. 
Finally, we perform a Borel transform with respect to the variable $p^2$. This leads to
\begin{equation}
    \hat F_\mu^{\, O_2}(p_D, q) = \frac{q_\mu }{m_{D}^2 f_{D} \pi} \int \limits_{m_c^2}^{s_0^D} ds\, e^{(m_{D}^2 - s)/M^2} \, {\rm Im}_s\big[F^{\, O_2}_{q}(s,q^2)\big]_{\rm OPE} \,,
    \label{eq:hatF-disp-int}
\end{equation}
where $M^2$ is the corresponding Borel parameter.
Proceeding in a similar way with the two-point correlator $\hat F_\mu^{\, O_2}(p_D, q)$, we can derive the corresponding dispersion relations in the $q^2$-channel. Inserting into eq.~\eqref{eq:two-point-correl} a complete set of states with the quantum number of the $\pi^-$ meson, yields
\begin{equation}
    \hat F_\mu^{\, O_2}(p_D, q) = \frac{i f_\pi q_\mu}{m_\pi^2 - q^2} \langle D(p_D) \pi(p_\pi)| O_2| \bar B(p_D+p_\pi)\rangle + q_\mu \! \int\limits_{s_{h^\prime}^{\prime\, (\pi)}}^\infty d s^\prime \,\, \frac{\rho_{h^\prime} (s^\prime)}{s^\prime - q^2} + \ldots \,,
    \label{eq:hatF-disp-rel-qsq}
\end{equation}
with $p_\pi^2 = m_\pi^2$ and $(p_D + p_\pi)^2 = m_B^2$. In eq.~\eqref{eq:hatF-disp-rel-qsq}, the pion decay constant is defined as $\langle 0| j_\mu^\pi| \pi(q)\rangle  = i f_\pi q_\mu$, while the spectral density $\rho_{h^\prime}(s^\prime)$ describes the contribution of excited states and of the continuum in the $q^2$-channel. 
Note that in writing the integral on the r.h.s.\ of eq.~\eqref{eq:hatF-disp-rel-qsq}, we have again taken into account that the correlation function $\hat F_\mu^{\, O_2}(p_D, q)$ admits the Lorentz decomposition analogous to the one in eq.~\eqref{eq:Fq-Fp}, however now with coefficients which can depend only on the variable $q^2$ since the first invariant is fixed. 
The matrix element we aim to determine is now on the r.h.s.\ of eq.~\eqref{eq:hatF-disp-rel-qsq}. 
Combining the latter with eq.~\eqref{eq:hatF-disp-int}, we obtain  
\begin{align}
 \frac{i f_\pi \langle O_2 \rangle}{m_\pi^2 - q^2}  & = \frac{1}{m_D^2 f_D \pi}  \int \limits_{m_c^2}^{s_0^D} ds\, e^{(m_{D}^2 - s)/M^2} \, {\rm Im}_s\big[F^{\, O_2}_{q}(s,q^2)\big]_{\rm OPE} - \int\limits_{s_{h^\prime}^{\prime\, (\pi)}}^\infty d s^\prime \,\, \frac{\rho_{h^\prime} (s^\prime)}{s^\prime - q^2}\,.
 \label{eq:ME-O1t-SR-no-QHD}
\end{align}
The matrix element $\langle O_2 \rangle$ could in principle be extracted by fitting the r.h.s.\ of eq.~\eqref{eq:ME-O1t-SR-no-QHD}. In this case, one could further isolate the next resonance due to the $a_1$-meson state and employ an ansatz, usually polynomial, to parametrise the remaining contribution due to the continuum.
However, this turns out to be practically not feasible, given that the current size of the theoretical uncertainties, strongly affected by the limited accuracy of many input parameters, see section~\ref{sec:Inputs}, makes the disentanglement of the pion state, of the $a_1$ state and of the continuum extremely challenging.
On the other hand, taking into account the approximate $1/q^2$ behaviour of
the OPE result in eq.~\eqref{eq:I-alphabetamu}, which almost perfectly matches the dominant contribution due to the pion pole on the l.h.s.\ of eq.~\eqref{eq:ME-O1t-SR-no-QHD}, one can already obtain 
a good estimate of the matrix element $\langle O_2 \rangle$, by considering only the first term on the r.h.s.\ of eq.~\eqref{eq:ME-O1t-SR-no-QHD}. 

Alternatively, expressing the OPE result on the r.h.s.\ of eq.~\eqref{eq:ME-O1t-SR-no-QHD} as a dispersive integral in the complex $q^2$-plane, with the first pole being on the real axis $ s^\prime =0$, and
recurring again to QHD, we can approximate
\begin{equation}
\int\limits_{s_{h^\prime}^{\prime\, (\pi)}}^\infty d s^\prime \,\, \frac{\rho_{h^\prime} (s^\prime)}{s^\prime - q^2} = 
\frac{1}{m_D^2 f_D \pi^2}  \int \limits_{s_0^{\pi}}^\infty d s^\prime \int \limits_{m_c^2}^{s_0^D} ds\, e^{(m_{D}^2 - s)/M^2} \, \frac{{\rm Im}_{s^\prime}{\rm Im}_s\big[F^{\, O_2}_{q}(s,s^\prime)\big]_{\rm OPE}}{s^\prime - q^2}  \,,
\label{eq:QHD-qsq}
\end{equation}
with $s_0^\pi$ denoting the effective threshold parameter in the $\pi$ channel.
From eqs.~\eqref{eq:ME-O1t-SR-no-QHD}, \eqref{eq:QHD-qsq}, after applying a Borel transform with respect to the variable $q^2$, we arrive at the following sum rule for the non-factorisable matrix element
\begin{align}
i \langle O_2 \rangle  = \frac{- e^{m^2_\pi/M^{\prime 2}}}{ f_\pi f_{D} m_{D}^2}  \int \limits_{m_c^2}^{s_0^D} ds \! \int \limits_0^\infty \!  d \omega_1 \sum_{\hat \psi}  \hat \psi(\omega_1) \sum_{n=1}^{3}  \frac{c_n^{\hat \psi}(\omega_1, 0)}{(n-1)!}  \, e^{(m_{D}^2 - s)/M^2} \delta_{s}^{(n-1)} \left( \tilde s(\omega_1,0)- s \right)\,,
\label{eq:ME-O1t-SR-result}
\end{align}
where $M^{\prime 2}$ denotes the corresponding Borel parameter in the $q^2$-channel and the expression for ${\rm Im}_{s^\prime}{\rm Im}_s\big[F^{\, O_2}_{q}(s,s^\prime)\big]_{\rm OPE}$ follows from using eq.~\eqref{eq:Im-delta}, with $\delta^{(n-1)}_x(f(x))$ indicating the $(n-1)$-derivative of the delta function with respect to the variable~$x$. Finally, note that also in this way, in consistency with what discussed below eq.~\eqref{eq:ME-O1t-SR-no-QHD},
because of the $1/q^2$ structure of the OPE result, only the contribution due to the pion pole enters eq.~\eqref{eq:ME-O1t-SR-result} and the sum rule becomes independent of $s_0^\pi$.
\section{Determination of \boldmath $\langle O_1 \rangle$ from LCSR}
\label{sec:ME-O1-QCDSR}
\begin{figure}[t]
    \centering
    \includegraphics[scale=0.5]{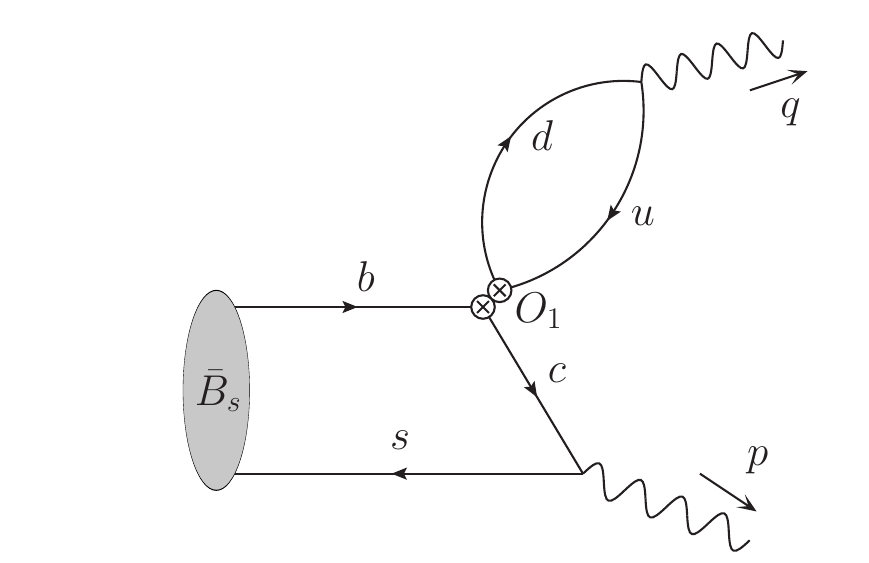}
    \includegraphics[scale=0.5]{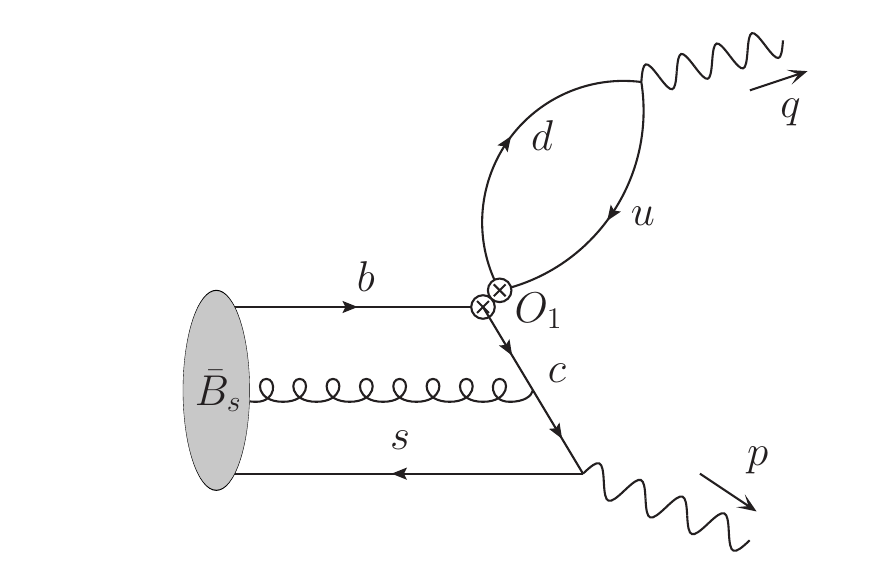}
    \caption{Diagrams describing the leading contributions due to two-particle (left) and three-particle~(right) LCDAs in the OPE for the correlator $F^{\, O_1}_\mu (p, q)$.}
    \label{fig:correlator-OPE-O1}
\end{figure}
The computation of the factorisable part of the amplitude $\langle O_1 \rangle$ within the framework of LCSR proceeds in a very similar manner to that discussed in the previous section. Therefore, here we limit ourselves to describing only the key steps. The starting point is now the following three point correlation function 
\begin{equation}
F^{\, O_1}_\mu (p, q) = i^2 \int d^4 x \, e^{i p \cdot x} \int d^4 y
\, e^{i q \cdot y} \,
\langle 0| {\rm T} \! \left\{j^{D}_5 (x), 
O_1 (0), j^\pi_\mu (y) \right\} 
| \bar B (p + q) \rangle\,,
\label{eq:Correlator-O1}
\end{equation}
where the two interpolating currents coincide with those in eq.~\eqref{eq:Correlator-O1T}. The kinematics is also chosen to be the same, i.e.\ $P^2 \equiv - p^2 \gg \Lambda^2$ and $Q^2 \equiv -q^2 \gg \Lambda^2$, so that the time-ordered product in eq.~\eqref{eq:Correlator-O1} is again calculated around $x^2 \sim 0$ and $y^\mu \sim 0$. Specifically, from eq.~\eqref{eq:Correlator-O1} we obtain
\begin{align}
F^{\, O_1}_\mu(p,q) & = - i N_c m_c  \int  d^4 x  \int d^4 y  \,\, e^{i p\cdot x} \, e^{i q \cdot y}  \,\, \langle 0| \bar s^i(x)\gamma_5 \, i S^{(c)}_{ij}(x,0) \gamma_\rho (1-\gamma_5) 
\nonumber \\[1mm]
& \times i S^{(u)}_{0}(-y) \gamma_\mu \gamma_5 \, i S_0^{(d)}(y) \gamma^\rho (1-\gamma_5) b^j (0)|\bar B(p+q)  \, \rangle \,,
\label{eq:Correlator-O1-2}
\end{align}
with $S^{(c)} (x,0)$ denoting the charm-quark propagator expanded near the light-cone. Including the leading one gluon corrections, this reads~\cite{Rusov:2017chr, Belyaev:1994zk}
\begin{align}
        S_{ij}^{(c)}(x,0) & =  
	- \frac{i m_c^2 \delta_{ij} }{4 \pi^2} \left[ \frac{K_1 (m_c \sqrt{-x^2})}{\sqrt{-x^2}} +
	i \frac{\slashed x}{-x^2} K_2(m_c \sqrt{-x^2}) \right]
 \nonumber \\
	& -  
	\frac{i \, t^a_{ij} }{16 \pi^2}   \int \limits_0^1 d u \Biggl[
	m_c \, K_0 (m_c \sqrt{-x^2}) G_{\mu \nu}^a(u x) \sigma^{\mu \nu} 
      \nonumber\\
	& +  
	\frac{i m_c}{\sqrt{-x^2}} K_1 (m_c \sqrt{-x^2})
	\left[ \bar u \slashed x G^a_{\mu \nu}(u x)  \sigma^{\mu \nu} + u G^a_{\mu \nu}(u x) \sigma^{\mu \nu} \slashed x \right] \Biggl] + \ldots \,,
 \label{eq:LC-charm-prop}
\end{align}
where the first line corresponds to the free-quark propagator already introduced in eq.~\eqref{eq:prop-coordinate-space}, and the ellipses indicate subleading corrections with at least one additional covariant derivative of the gluon field strength tensor; note also that in writing eq.~\eqref{eq:Correlator-O1-2} we have already taken into account that the colour structure now forbids the emission of one gluon from the light-quark loop and we have thus replaced the two propagators with the corresponding free quark ones, see Figure~\ref{fig:correlator-OPE-O1}.

The integral over $y^\mu$ in eq.~\eqref{eq:Correlator-O1-2} can be easily performed. In dimensional regularisation it yields the standard massless one-loop two-point function,
and, as expected, the result is transversal with respect to the momentum of the light-quark current $q^\mu$. On the other hand, the integration over $x^\mu$ can be computed once a parametrisation for the corresponding two- and three-particle $B$-meson matrix elements is implemented. Using the results given in appendix~\ref{app:B}, again in the HQET limit, we have respectively
\begin{align}
 \langle 0| \bar s_\alpha (x) b_{\beta}(0)|\bar B(p+q) \rangle & = - F_{B}(\mu) \sqrt{m_{B}}  \int_0^\infty \!\! d \omega  \,  e^{- i \omega v \cdot x} \Bigg\{\frac{i}{2}\big(\phi_+ + x^2 g_+) P_+ \gamma_5 
\nonumber \\
& + \frac{1}{4} \left[(\bar \phi_+ - \bar \phi_-) + x^2 (\bar g_+ - \bar g_-)\right] P_+ \, \slashed x \, \gamma_5 \Bigg\}_{\beta \alpha}\!\!\!(\omega ; \mu)\,,
\label{eq:2-part-ME-fac}
\end{align}
and
\begin{align}
    & \langle 0| \bar s_\alpha (x) G_{\mu \nu}(ux) b_{\beta}(0)|\bar B(p+q) \rangle = \frac12 F_{B}(\mu) \sqrt{m_{B}}
    \int_0^\infty \!\! d \omega_1  \int_0^\infty \!\! d \omega_2 \,  e^{- i (\omega_1+ u \omega_2) v \cdot x}
    \nonumber \\[2mm]
    & 
    \qquad \times
    \Big\{P_+ \big[ - i \sigma_{\mu \nu} \psi_V
    + (v_\mu \gamma_\nu - v_\nu \gamma_\mu)(\psi_A - \psi_V)   
    - i (x_\mu v_\nu - x_\nu v_\mu) \bar{\psi}_{X_A} 
    \nonumber \\[2mm]
    &
    \qquad \qquad + i (x_\mu \gamma_\nu - x_\nu \gamma_\mu) (\bar{\psi}_W + \bar{\psi}_{Y_A}) - \epsilon_{\mu \nu \eta \tau} x^\eta v^\tau \gamma_5 \bar{\psi}_{\tilde{X}_A}  + \epsilon_{\mu \nu \eta \tau} x^\eta \gamma^\tau \gamma_5  \bar{\psi}_{\tilde{Y}_A}
    \nonumber \\[2mm]
    &
    \qquad \qquad + (x_\mu v_\nu - x_\nu v_\mu) \slashed x \bar{\bar{\psi}}_W - (x_\mu \gamma_\nu - x_\nu \gamma_\mu) \slashed x \bar{\bar{\psi}}_ Z
    \big] \gamma_5 \Big\}_{\beta \alpha}\!\!(\omega_1, \omega_2;\mu) \,,
    \label{eq:3-part-ME-fac}
\end{align}
with the notation introduced in eqs.~\eqref{eq:Psi-bar-def}, \eqref{eq:phi-bar-app}.  Substituting eq.~\eqref{eq:LC-charm-prop} into eq.~\eqref{eq:Correlator-O1-2} and using eqs.~\eqref{eq:2-part-ME-fac}, \eqref{eq:3-part-ME-fac}, we are left with the evaluation of the same type of tensor integrals as those in eqs.~\eqref{eq:int-x-K1}, \eqref{eq:int-x-K2}, together with the following one
\begin{equation}
\int d^4 x \, e^{i \tilde p \cdot x}  \, K_0(m_c \sqrt{-x^2}) \,  \Big\{1, x^\mu, x^\mu x^\nu, \ldots  \Big\} \,,
\label{eq:int-x-K0}
\end{equation}
with $\tilde p^\mu = p^\mu - \omega v^\mu$, and $\tilde p^\mu = p^\mu - (\omega_1 + u \, \omega_2) v^\mu$, respectively for the two- and three-particle contributions.
Using the expressions for the inverse Fourier transforms of Bessel functions collected in appendix~\ref{app:D}, we arrive at the final form of the three-point correlator in eq.~\eqref{eq:Correlator-O1}, that is
\begin{equation}
 F^{\, O_1}_\mu(p,q) = F^{\,  O_1}_q (p^2,q^2) \, q_\mu +
F^{\, O_1}_p (p^2,q^2) \, p_\mu\,,
    \label{eq:F-O1-transverse}
\end{equation}
where the contributions to the invariant amplitude $F^{\,  O_1}_q (p^2,q^2)$ due to the two- and three-particle matrix elements are written in terms of a LC OPE, respectively, as 
\begin{equation}
\left[F^{\,  O_1}_q (p^2,q^2)\right]_{\rm OPE, 2p} =  F_B \sqrt{m_B} \, m_c \int\limits_0^\infty d \omega \sum_\phi \phi(\omega) \sum_{n= 1}^{4} \frac{c^\phi_n (\omega, q^2)}{[\tilde s(\omega, q^2) - p^2 - i \varepsilon]^n} \ln\left(- \frac{q^2}{\mu^2}\right) \,,
\label{eq:OPE-fac-2p}
\end{equation}
with $\phi = \phi_+, \bar g_+, \ldots $, and
\begin{align}
\big[F^{\,  O_1}_q (p^2,q^2)]_{\rm OPE, 3p} &=  F_B \sqrt{m_B} \, m_c \int \limits_0^1 d u  \int \limits_0^\infty d \omega_2 \int \limits_{u \omega_2}^{\infty} \!\! d \omega \sum_\psi \psi(u, \omega_2,\omega) 
\nonumber \\
& \times \sum_{n=1}^{4} \frac{c^\psi_n (u, \omega, q^2)}{[\tilde s(\omega, q^2) - p^2 - i \varepsilon]^n} \ln\left(- \frac{q^2}{\mu^2}\right),
\label{eq:OPE-fac-3p}
\end{align}
with $\psi = \psi_A, \psi_V, \ldots$. The function $\tilde s (\omega, q^2)$ in eqs.~\eqref{eq:OPE-fac-2p}, \eqref{eq:OPE-fac-3p}, is defined as in eq.~\eqref{eq:tilde-s-func}, while the analytic expressions of the OPE coefficients $c_n^\phi(\omega, q^2)$, $c_n^\psi(u,\omega, q^2)$ can be found in appendix~\ref{app:E}. Note that both the divergent $1/\epsilon$ piece and the remaining constant term originating from the light-quark loop have been omitted, as only the coefficient proportional to $\log(-q^2/\mu^2)$ is relevant for the derivation of the dispersion relations. To this end, we follow the same procedure as done in the previous section, employing QHD as well as applying a Borel transform in both the $p^2$- and $q^2$-channels.  The final result can be compactly presented as
\begin{equation}
    i \langle O_1 \rangle = \frac{1}{\pi^2 f_\pi f_D \, m_D^2} \int\limits_0^{s_0^\pi} d s^\prime \!  \int\limits_{m_c^2}^{s_0^D} d s \,\,\,  e^{(m_{D}^2 - s)/M^2} \,  e^{(m_{\pi}^2 - s^\prime)/M^{\prime 2}}\,  {\rm Im}_{s^\prime}{\rm Im}_s\big[F^{\, O_1}_{q}(s,s^\prime)\big]_{\rm OPE} \,,
    \label{eq:O1-SR}
\end{equation}
where $\big[F^{\, O_1}_{q}(s,s^\prime)\big]_{\rm OPE}$ includes both the two- and three-particle contributions given in eqs.~\eqref{eq:OPE-fac-2p}, \eqref{eq:OPE-fac-3p}, and the corresponding imaginary part can be easily obtained from the identities given in eqs.~\eqref{eq:Im-delta}, \eqref{eq:Im-log}.
\section{Numerical analysis}
\label{sec:Num-analysis}
\subsection{Discussion of the inputs}
\label{sec:Inputs}
Below we discuss the numerical value of the inputs used in our analysis \footnote{In this section the notation $B^0$ and $B_d$ is used interchangeably.}. 
Following Ref.~\cite{Braun:2017liq}, the eight LCDAs, arising in the parametrisation of the three-particle $B$-meson matrix element in eq.~\eqref{eq:app:3-Part-ME}, are decomposed in terms of DAs of definite collinear twist, see eq.~\eqref{eq:LCDAs-twist}. 
These non-perturbative inputs can then be estimated by constructing specific model-dependent parametrisations, all satisfying the same normalisation conditions and asymptotic behaviour for small value of the arguments~\cite{Braun:2017liq}.
In our analysis, we employ the exponential model. Specifically, we follow Refs.~\cite{Braun:2017liq, Khodjamirian:2006st} for the twist-3 and twist-4 LCDAs and use, respectively
\begin{eqnarray}
\phi_3 (\omega_1, \omega_2) & = &
\frac{\lambda_E^2 - \lambda^2_H}{6 \, \omega_0^5} \omega_1 \omega_2^2\, e^{-(\omega_1+\omega_2)/\omega_0}\,,
\label{eq:phi-3-EM}
\\
\phi_4 (\omega_1, \omega_2) & = &
\frac{\lambda_E^2 + \lambda^2_H}{6 \, \omega_0^4} \omega_2^2 \, e^{-(\omega_1+\omega_2)/\omega_0}\,,
\label{eq:phi-4-EM}
\\
\psi_4 (\omega_1, \omega_2) & = &
\frac{\lambda_E^2}{3 \, \omega_0^4} \omega_1 \omega_2 \, e^{-(\omega_1+\omega_2)/\omega_0}\,, 
\label{eq:psi-4-EM}
\\
\tilde \psi_4 (\omega_1, \omega_2) & = &
\frac{\lambda^2_H}{3 \, \omega_0^4} \omega_1 \omega_2 \, e^{-(\omega_1+\omega_2)/\omega_0}\,,
\label{eq:psi-4-t-EM}
\end{eqnarray}
whereas for the twist-5 and twist-6 LCDAs we use the parametrisation proposed in Ref.~\cite{Lu:2018cfc}, namely
\begin{eqnarray}
\tilde \phi_5 (\omega_1, \omega_2) & = &
\frac{\lambda_E^2 + \lambda^2_H}{3 \, \omega_0^3} \omega_1 \, e^{-(\omega_1+\omega_2)/\omega_0}\,,
\label{eq:phi-5-EM}
\\
\psi_5 (\omega_1, \omega_2) & = &
- \frac{\lambda_E^2}{3 \, \omega_0^3} \omega_2 \, e^{-(\omega_1+\omega_2)/\omega_0}\,,
\label{eq:psi-5-EM}
\\
\tilde \psi_5 (\omega_1, \omega_2) & = &
- \frac{\lambda_H^2}{3 \, \omega_0^3} \omega_2 \, e^{-(\omega_1+\omega_2)/\omega_0}\,, 
\label{eq:psi-5-t-EM}
\\
\phi_6 (\omega_1, \omega_2) & = &
\frac{\lambda^2_E - \lambda_H^2}{3 \, \omega_0^2} \, e^{-(\omega_1+\omega_2)/\omega_0}\,.
\label{eq:phi-6-EM}
\end{eqnarray}
In the studies performed e.g.\ in Refs.~\cite{Gubernari:2018wyi, Gubernari:2020eft}, the expansion has been truncated at twist-4 so that the DAs in eqs.~\eqref{eq:phi-5-EM}-\eqref{eq:phi-6-EM} were neglected. In fact, the LCDAs of twist-5 and twist-6 were not expected to contribute at the current accuracy of ${\cal O}(1/m_B)$~\cite{Braun:2017liq}, and in addition, the four DAs in eqs.~\eqref{eq:phi-5-EM}-\eqref{eq:phi-6-EM} would not be exhaustive for a complete description of the three-particle matrix element up to twist-6, since other LCDAs of the same order would still be missing~\cite{Gubernari:2018wyi}. However, we stress that the inclusion of the twist-5 and twist-6 DAs in eqs.~\eqref{eq:phi-5-EM}-\eqref{eq:phi-6-EM} is actually necessary to ensure that eq.~\eqref{eq:app:3-Part-ME} has the correct local limit \footnote{From the local limit of eq.~\eqref{eq:app:3-Part-ME}, cf.\ eq.~(5.1) of Ref.~\cite{Braun:2017liq}, it follows that $\Psi_V(0,0)=(1/3)\lambda_H^2$, $\Psi_A(0,0)=(1/3)\lambda_E^2$, and $\Psi_{X_A}(0,0)= \ldots = \Psi_Z(0,0) = 0 $. However, truncating at twist-4, i.e.\ neglecting $\tilde \Phi_5, \ldots, \Phi_6$, in eq.~\eqref{eq:LCDAs-twist}, leads instead to $\Psi_{Y_A}(0,0) = (-1/6) \lambda_E^2$ and $\Psi_{\tilde Y_A}(0,0) = (1/6) \lambda_H^2 $.},
and therefore we refrain from truncating the expansion at twist-4.  
Moreover, as discussed in the next section and as shown in Table~\ref{tab:part-contrib-LCDAs}, we find that neglecting these higher-twist DAs leads to pronounced cancellations, mainly because, in this case, the contribution due to $\psi_{\tilde Y_A}$ is found to largely compensate the one due to $\psi_V$. On the other hand, when including also the twist-5 and twist-6 LCDAs, the coefficient of $\psi_{\tilde Y_A}$
becomes roughly one order of magnitude smaller and no cancellations between LCDAs arise.

Turning to the two-particle DAs, we again adopt the exponential model \footnote{Several different models, mostly for the twist-2 LCDA $\phi_+$, have been proposed and studied in the recent literature, see e.g.\ Refs.~\cite{Braun:2017liq, Beneke:2018wjp, Feldmann:2022uok}.} and use, for the LCDAs up to twist-4, the parametrisation given in Ref.~\cite{Braun:2017liq}, i.e.
\begin{eqnarray}
\phi_+ (\omega) & = & 
\frac{\omega}{\omega_0^2} e^{-\omega/\omega_0}\,, 
\label{eq:phi-p-EM} \\
\phi_- (\omega) & = & 
\frac{e^{-\omega/\omega_0}}{\omega_0} - \frac{\lambda_E^2 - \lambda_H^2}{9 \, \omega_0^3} e^{-\omega/\omega_0} \left[1 - 2 \frac{\omega}{\omega_0} + \frac{1}{2} \frac{\omega^2}{\omega_0^2} \right],
\label{eq:phi-m-EM} \\
 g_+ (\omega) & = & 
 \frac{\omega^2}{2 \omega_0} \left(1 - \frac{\lambda_E^2 - \lambda_H^2}{36 \, \omega_0^2} \right) e^{-\omega/\omega_0}
 - \frac{\lambda_E^2}{6 \omega_0^2} \Biggl[\left(\omega - 2 \, \omega_0 \right) {\rm Ei}\left(-\frac{\omega}{\omega_0}\right) 
  \nonumber \\
 & & 
 + \left(\omega + 2 \omega_0 \right) e^{-\omega/\omega_0} \left(\ln \frac{\omega}{\omega_0} + \gamma_E \right) - 2 \omega \, e^{-\omega/\omega_0} \Biggr],
\label{eq:g-p-EM}
\end{eqnarray}
where ${\rm Ei(z)}$ is the exponential integral and $\gamma_E$ is the Euler constant, while for the twist-5 LCDA we follow Ref.~\cite{Lu:2018cfc} and use 
\begin{equation}
g_- (\omega) = \omega \left[\frac{3}{4} - \frac{\lambda_E^2-\lambda_H^2}{12 \, \omega_0^2} \left(1 - \frac{\omega}{\omega_0} + \frac{1}{3} \frac{\omega^2}{\omega_0^2}\right)\right] e^{-\omega/\omega_0}\,.
\label{eq:g-m-EM}
\end{equation}
The models in eqs.~\eqref{eq:phi-3-EM}-\eqref{eq:g-m-EM} depend on the parameters $\omega_0$, $\lambda_E^2$, and $\lambda_H^2$. Within the exponential model and using EOM relations, it follows that $\omega_0 = \lambda_B$~\cite{Braun:2017liq}, with $\lambda_B$ being the inverse moment of the two-particle $B$-meson distribution amplitude $\phi_+ (\omega)$. The remaining two parameters $\lambda_E^2$ and $\lambda_H^2$ characterise the local vacuum-to-$B$-meson quark-gluon-quark matrix element. These inputs must be determined with some non-perturbative techniques, and are currently still quite poorly known. Specifically, for the parameter $\lambda_B$, there exist several determinations in the literature, obtained either with QCD sum rules \cite{Braun:2003wx, Khodjamirian:2020hob}, OPE-based methods \cite{Lee:2005gza, Kawamura:2008vq, Kawamura:2010tj}\footnote{Very recently, a study of the strange quark mass effects has been preformed in Ref.~\cite{Feldmann:2023aml}.}, or from studies of the $B \to \gamma \ell \bar \nu$ decay \cite{Ball:2003fq, Beneke:2011nf, Braun:2012kp, Beneke:2018wjp}.
In our analysis, we use the recent sum rule result from Ref.~\cite{Khodjamirian:2020hob} where, for the first time, the complete SU(3)$_F$ breaking effects due to the strange quark mass
have been taken into account, hence providing estimates of the parameter $\lambda_B$ for both the $B$ mesons, i.e.\
\begin{align}
\lambda_{B_d}(1 \GeV) &= (0.383 \pm 0.153) \GeV\,, 
\\[2mm]
\lambda_{B_s}(1 \GeV) &= (0.438 \pm 0.150) \GeV\,.
\end{align}
As for the parameters $\lambda_E^2$ and $\lambda_H^2$, in the case of the $B_d$ meson, several studies within the framework of QCD sum rule have been performed \cite{Grozin:1996pq, Nishikawa:2011qk, Rahimi:2020zzo}. The first estimates, obtained in Ref.~\cite{Grozin:1996pq}, included only LO-QCD contributions up to dimension-five in the corresponding OPE, yielding respectively $\lambda_{E,B_d}^2(1 \GeV) = (0.11 \pm 0.06) \GeV^2 $ and $\lambda_{H,B_d}^2(1 \GeV) = (0.18 \pm 0.07) \GeV^2$.
Later, perturbative QCD corrections to the dimension-five contribution, as well as the LO-QCD dimension-six contributions were taken into account in Ref.~\cite{Nishikawa:2011qk}. These corrections improved the overall stability of the sum rule, leading to the smaller values $\lambda_{E, B_d}^2 (1 \GeV) = (0.03 \pm 0.02) \GeVsq$ and $\lambda_{H, B_d}^2 (1 \GeV)= (0.06 \pm 0.03) \GeVsq$. Recently, a new study, performed using a different expression for the correlation functions and including dimension-seven contributions,
has been carried out in Ref.~\cite{Rahimi:2020zzo}. The authors 
have obtained the values $\lambda_{E, B_d}^2 (1 \GeV) = (0.01 \pm 0.01) \GeVsq$ and $\lambda_{H,B_d}^2 (1 \GeV) = (0.15 \pm 0.05) \GeVsq$, where the former is consistent with the result of Ref.~\cite{Nishikawa:2011qk}, while the latter is considerably above. Therefore, to account for the spread in the two determinations,
in our analysis we use the following intervals
\begin{align}
\lambda_{E,B_d}^2 (1 \GeV) & = (0.03 \pm 0.03) \GeVsq\,,  
\\[2mm]
\lambda_{H, B_d}^2 (1 \GeV) & = (0.12 \pm 0.09) \GeVsq\,,
\end{align}
which cover the results of both Refs.~\cite{Nishikawa:2011qk, Rahimi:2020zzo}. On the other hand, since there are still no estimates of the parameters $\lambda_{E,B_s}^2$ and $\lambda_{H,B_s}^2$ available in the literature, we fix their central values to be the same as the corresponding ones for the $B_d$ meson,  
adding an extra $20 \%$ uncertainty to account for SU(3)$_F$ breaking effects. This gives
\begin{align}
\lambda_{E, B_s}^2 (1 \GeV) &= (0.03 \pm 0.04) \GeVsq\,,
\\[2mm]
\lambda_{H, B_s}^2 (1 \GeV) &= (0.12 \pm 0.11) \GeVsq\,.
\end{align}
Another important ingredient of the computation is the choice of the sum rule parameters. For the threshold continuum $s_0^D$ and the Borel parameter $M^2$ in the $D_{(s)}$-meson channel, we adopt the same intervals as used in the recent QCD sum rule studies of the form factors for the $B \to D$~\cite{Gubernari:2018wyi} and $B_s \to D_s$~\cite{Bordone:2019guc} transitions, see also Refs.~\cite{Khodjamirian:2005ea, Khodjamirian:2006st, Faller:2008tr}. We thus use respectively
\begin{align} 
    & s_0^{D^+} = (6.8 \pm 1.0) \GeVsq\,,
    & M_{D^+}^2 = (3 \pm 1.5) \GeVsq\,,
    \\
    & s_0^{D_s^+} = (9.0 \pm 2.1) \GeVsq\,,
    & M_{D_s^+}^2 = (3 \pm 1.5) \GeVsq\,,
\end{align}
while, for the corresponding sum rule parameters in the $\pi$- and $K$-meson channels, we use the following values \cite{Khodjamirian:2006st, Shifman:1978by, Colangelo:2000dp, Khodjamirian:2003xk}   
\begin{align}
    & s_0^{\pi^-} = (0.7 \pm 0.1) \GeVsq\,,  
    & M_{\pi^-}^2 = (1.0 \pm 0.5) \GeVsq\,, \\
    & s_0^{K^-} = (1.05 \pm 0.10) \GeVsq\,, 
    & M_{K^-}^2 = (1.0 \pm 0.5) \GeVsq\,.
\end{align}
\begin{table}[t]
\renewcommand{\arraystretch}{1.4}
    \centering
    \begin{tabular}{|c|c|c||c|c|c|}
     \hline
     \hline
     $m_{B^0}$ & $5.27965 \GeV$ & \cite{Workman:2022ynf} &
     $m_{B_s^0}$ & $5.36688 \GeV$ & \cite{Workman:2022ynf} \\
     $m_{D^+}$ & $1.86965 \GeV$ & \cite{Workman:2022ynf} &
     $m_{D_s^+}$ & $1.96834 \GeV$ & \cite{Workman:2022ynf} \\
     $m_{K^+}$ & $0.493677 \GeV$ & \cite{Workman:2022ynf} &
     $m_{\pi^+}$ & $0.13957 \GeV$ & \cite{Workman:2022ynf} \\
     \hline
     $\tau_{B^0}$ & $(1.519 \pm 0.004)$~ps &
     \cite{Workman:2022ynf} & 
     $\tau_{B_s^0}$ & $(1.527 \pm 0.011)$~ps &\cite{Workman:2022ynf} \\
     \hline 
     $f_{B_d}$ & $0.1900 \GeV$ & \cite{FLAG:2019iem} &
     $f_{B_s}$ & $0.2303 \GeV$ & \cite{FLAG:2019iem} \\
     $f_{D^+}$ & $0.2120 \GeV$ & \cite{FLAG:2019iem} &
     $f_{D_s^+}$ & $0.2499 \GeV$ & \cite{FLAG:2019iem} \\
     $f_{K^+}$   & $0.1556 \GeV$ & \cite{FLAG:2019iem} &
     $f_{\pi^+}$ & $0.1302 \GeV$ & \cite{FLAG:2019iem} \\
     \hline
     $\lambda_{B_d} $ & $(0.383 \pm 0.150) \GeV$ & \cite{Khodjamirian:2020hob} &
     $\lambda_{B_s} $ & $(0.438 \pm 0.150) \GeV$ & \cite{Khodjamirian:2020hob} \\
     $\lambda_{E,B_d}^2 $ & $(0.03 \pm 0.03) \GeVsq$ & 
     \cite{Nishikawa:2011qk, Rahimi:2020zzo} &
     $\lambda_{E,B_s}^2 $ & $(0.03 \pm 0.04) \GeVsq$ & ${\rm SU}(3)_F$ \\
     $\lambda_{H,B_d}^2 $ & $(0.12 \pm 0.09) \GeVsq$ & \cite{Nishikawa:2011qk, Rahimi:2020zzo} &
     $\lambda_{H,B_s}^2 $ & $(0.12 \pm 0.11) \GeVsq$ & ${\rm SU}(3)_F$ \\
     \hline
     $M_{D^+}^2$ & $(4.5 \pm 1.5) \GeVsq$ & \cite{Gubernari:2018wyi} &
     $M_{D^+_s}^2$ & $(4.5 \pm 1.5) \GeVsq$ & \cite{Bordone:2019guc} \\
     $s_0^{D^+}$ & $(6.8 \pm 1.0) \GeVsq$ & \cite{Gubernari:2018wyi} &
     $s_0^{D_s^+}$ & $(9.0 \pm 2.1) \GeVsq$ & \cite{Bordone:2019guc} \\
     \hline
     $M_{K^-}^{2}$ & $(1.0 \pm 0.5) \GeVsq$ & \cite{Khodjamirian:2006st} &
     $M_{\pi^-}^{2}$ & $(1.0 \pm 0.5) \GeVsq$ & \cite{Khodjamirian:2006st} \\
     $s_0^{K^-}$ & $(1.05 \pm 0.1) \GeVsq$ & \cite{Khodjamirian:2003xk} & 
     $s_0^{\pi^-}$ & $(0.7 \pm 0.1) \GeVsq$ & \cite{Khodjamirian:2003xk}  \\
     \hline
     \hline
     $|V_{ud}|$ 
     & $0.97435 \pm 0.00016$  
     & \cite{Workman:2022ynf}
     & $|V_{us}|$ 
     & $0.22500 \pm 0.00067$ 
     & \cite{Workman:2022ynf} 
     \\
     $|V_{cb}|$
     & $0.04182^{+0.00085}_{-0.00074}$ 
     & \cite{Workman:2022ynf}
     & $\alpha_s (M_Z) $ 
     & $0.1179 \pm 0.0009$ 
     & \cite{Workman:2022ynf}  
     \\
     $\overline m_b (\overline m_b)$
     & $(4.18 \pm 0.03) \GeV $ 
     & \cite{Workman:2022ynf}
     & $ \overline m_c (\overline m_c)$  
     & $(1.27 \pm 0.02) \GeV$ 
     & \cite{Workman:2022ynf}
     \\
     \hline 
     \hline
    \end{tabular}
    \caption{Summary of the inputs used in the numerical analysis.
    The values of the parameters $\lambda_B$, $\lambda_{E}^2$, and $\lambda_{H}^2$, correspond to $\mu = 1 \GeV$.}
    \label{tab:num-input}
\end{table}
The QCD decay constants are determined with high precision within Latice QCD, and for all the mesons considered we take the corresponding FLAG values~\cite{FLAG:2019iem}.
As for the HQET decay constant $F_B (\mu)$, which enters eq.~\eqref{eq:3-part-ME}, we use the one-loop relation to the QCD decay constant $f_B$, valid up to power corrections of the order of $1/m_b$ \cite{Neubert:1991sp}, namely
\begin{equation}
F_B (\mu) = f_B \sqrt{m_B} \left[1 - \frac{C_F \, \alpha_s(\mu)}{4 \pi} 
\left(3 \, \ln \frac{m_b}{\mu} -2 \right)\right] + \ldots\,,
\label{eq:FB-mu}
\end{equation}
with $C_F = 4/3$. In our analysis, the central value of the renormalisation scale in eq.~\eqref{eq:FB-mu} is set to $\mu =1$ GeV, corresponding to the scale at which the inputs $\lambda_B, \lambda_{E}^2$, and $\lambda_H^2$, have been determined. Taking then into account the scale-dependence of the latter parameters \cite{Braun:2003wx, Nishikawa:2011qk, Grozin:1996hk}, the total uncertainty due to $\mu$-variation is obtained varying this scale in the interval $1 \GeV \le \mu \le 1.5 \GeV$.
For the strong coupling $\alpha_s (\mu)$, we include the five-loop running implemented in the Mathematica package $\mathtt{ RunDec}$~\cite{Herren:2017osy}  
and use the most recent result \cite{Workman:2022ynf}
\begin{equation*}
\alpha_s (M_Z) = 0.1179 \pm 0.0009\,.   
\end{equation*}
For the quark masses, we use the corresponding values in the $\overline{\rm MS}$-scheme, i.e.\ $\overline m_b (\overline m_b) = (4.18 \pm 0.03) \GeV$ and
$\overline m_c (\overline m_c) = (1.27 \pm 0.02) \GeV$~\cite{Workman:2022ynf}. Values of the meson masses, known very precisely, are also taken from the PDG~\cite{Workman:2022ynf}.

In order to obtain predictions for the branching fractions, we need in addition to fix the value of the Wilson coefficients, of the CKM matrix elements, and of the $ B_{(s)}^0$-meson lifetime. For the former, we use the corresponding results at NLO accuracy, see e.g.\ Ref.~\cite{Gorbahn:2004my}. The central value of the Wilson coefficients is obtained setting $\mu_b = \overline m_b$,
and this scale is then varied in the interval $\overline m_b/2 \leq \mu_b \leq 2 \overline m_b$. We stress that the choice of using NLO results, despite the LO accuracy of the corresponding matrix elements, is motivated by the fact that there is a sizeable shift of $\sim -40 \%$ in the value of $C_2$, when going from LO to NLO \footnote{The shift from NLO to NNLO can be instead neglected, given the current accuracy of our study.}, which strongly affects the prediction of the non-factorisable part of the amplitude. The computation of the missing perturbative QCD corrections to the matrix elements would be clearly of utmost importance in order to assess the total size of NLO effects.

For the CKM matrix elements, we use the best-fit values, obtained from a global fit, 
provided by the PDG \cite{Workman:2022ynf}, i.e.\
\begin{equation*}
|V_{ud}| = 0.97435 \pm 0.00016, \quad
|V_{us}| = 0.22500 \pm 0.00067, \quad
|V_{cb}| = 0.04182^{+0.00085}_{-0.00074}\,.
\end{equation*}
Finally, $B_{(s)}^0$-meson lifetimes are by now measured very precisely and their values are taken from Ref.~\cite{Workman:2022ynf}.\footnote{$B_{(s)}^0$-meson lifetimes can also be computed using the framework of the Heavy Quark Expansion. However, the current theoretical uncertainties are still much larger than the corresponding experimental ones~\cite{Lenz:2022rbq}.
}
For convenience, all the inputs used in our analysis are collected in Table~\ref{tab:num-input}.
\subsection{Results}
\label{sec:Results}
In this section we present our predictions, obtained within the framework of LCSR and at LO-QCD, of the factorisable and non-factorisable matrix elements for the non-leptonic decays $\bar B_s^0 \to D_s^+ \pi^-$ and $\bar B^0 \to D^+ K^-$, as well as of the corresponding branching fractions.

\begin{figure}[th]
    \centering
    \includegraphics[scale=0.75]{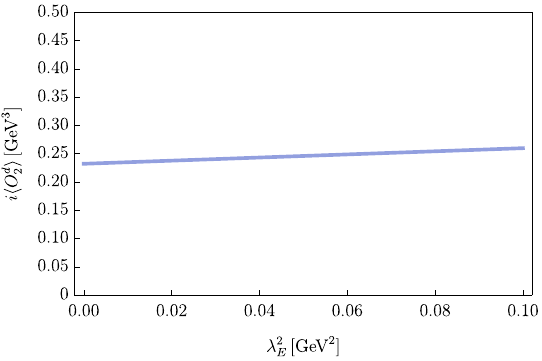}
    \includegraphics[scale=0.75]{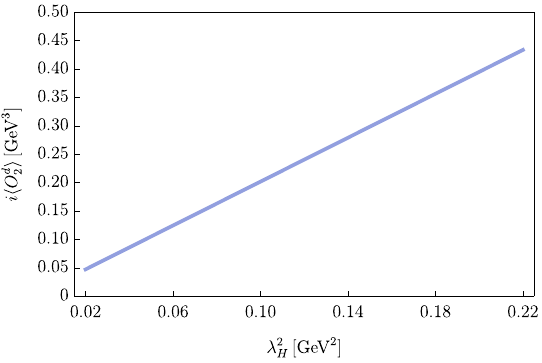} \\[2mm]
    \includegraphics[scale=0.75]{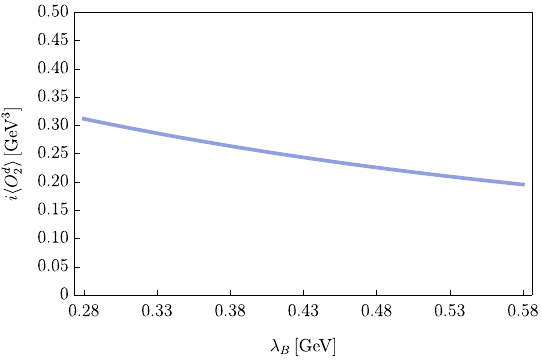}
    \includegraphics[scale=0.75]{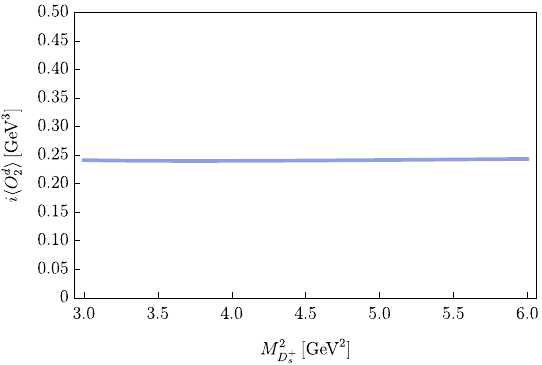} \\[2mm]
    \includegraphics[scale=0.75]{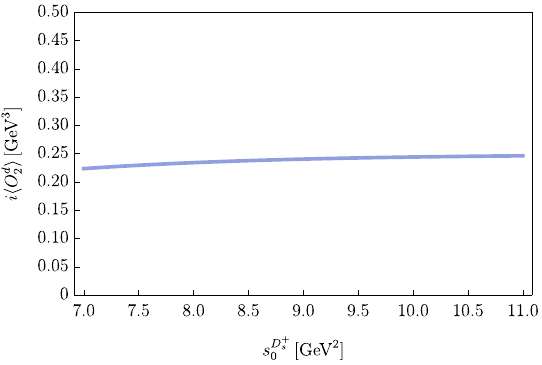}
    \includegraphics[scale=0.75]{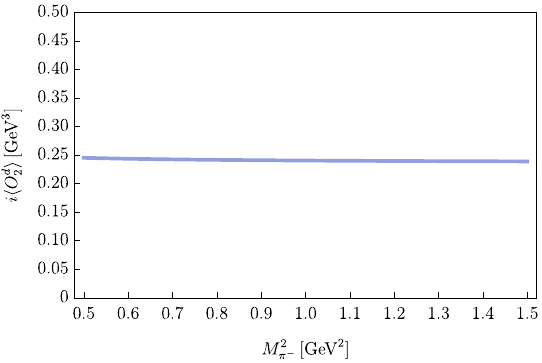} \\[2mm]
    \includegraphics[scale=0.75]{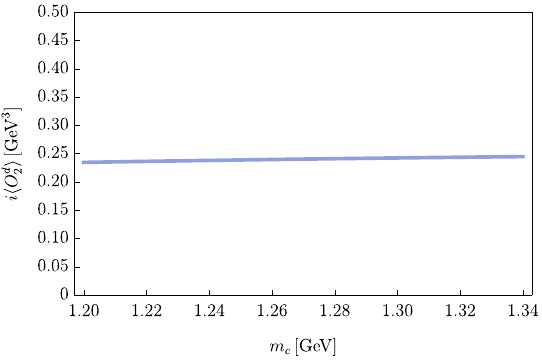} 
    \includegraphics[scale=0.75]{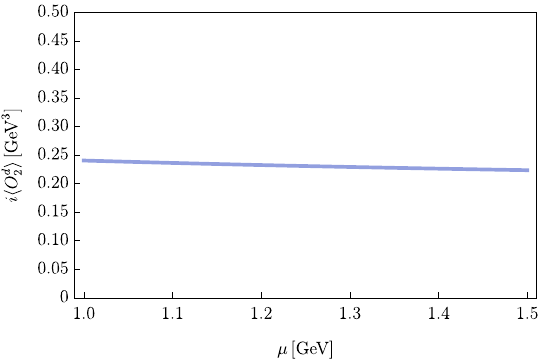} 
    \caption{Dependence of the non-factorisable matrix element $i \langle O_2^d \rangle$ on the LCDAs inputs $\lambda_E^2, \lambda_H^2, \lambda_B$, the sum rule parameters $M_{D^+}^2, s_0^D, M^2_{\pi^-}$, as well as on the charm-quark mass $m_c$ and the renormalisation scale $\mu$. In each plot, 
    the remaining parameters are fixed to their central values.}
    \label{fig:par-depend-ME01}
\end{figure}

Let us start by discussing the predictions for the non-factorisable matrix element~$\langle O^q_{2} \rangle$, which represents the main result of the paper. The final sum rule is given in eq.~\eqref{eq:ME-O1t-SR-result}, and in order to illustrate the main sources of uncertainty, in Figure~\ref{fig:par-depend-ME01} we show the dependence of $i \langle O_2^d \rangle$ on different inputs, fixing in each plot the remaining parameters to their central values. For easier comparison, all plots are displayed in the same interval, namely $i \langle O_2^d \rangle \in [0,0.50] \, {\rm GeV}^3$, and for brevity, we only show the mode $\bar B_s^0 \to D_s^+ \pi^-$, since the behaviour of the corresponding matrix element in the case of the $\bar B^0 \to D^+ K^-$ decay is completely analogous. We find that the sum rule prediction for the non-factorisable matrix element is extremely sensitive to the value of the parameter $\lambda_H^2$ which, on the other hand, as discussed in the previous section, is still poorly known. The result is also quite sensitive to the size of $\lambda_B$, while the dependence on $\lambda_E^2$ appears softer.
Clearly, a more precise determination of these non-perturbative inputs is essential in order to improve the accuracy of the present analysis. The sensitivity to the value of the threshold continuum $s_0^D$, and of the Borel parameters $M_D^2$, and $M^2_{\pi}$, is found to be quite mild, thus reflecting the overall stability of the sum rule. The dependence on the charm quark mass and on the renormalisation scale $\mu$ is also very moderate. 

The partial contribution to $i \langle O_2^d \rangle$, for each of the eight LCDAs entering the parametrisation of the three-particle $B$-meson matrix element in eq.~\eqref{eq:app:3-Part-ME}, is shown in the third column of Table~\ref{tab:part-contrib-LCDAs}, in correspondence of the central values of all the input parameters. 
We find that the function $\psi_V$ gives the dominant contribution to the non-factorisable matrix element, while the remaining LCDAs lead all together to a small effect. As stated in the previous section, in our analysis we
use the results for the LCDAs up to twist-six accuracy; however, 
for comparison, in the last column of Table~\ref{tab:part-contrib-LCDAs}, we also provide the corresponding partial contributions to $i \langle O^d_2 \rangle$ obtained neglecting the twist-5 and twist-6 LCDAs in eqs.~\eqref{eq:phi-5-EM}-\eqref{eq:phi-6-EM}.
In this case, there is a strong cancellation between the coefficients of $\psi_V$ and $\psi_{\tilde Y_A}$, leading to a much smaller value for $\langle O_2^d \rangle$. Again, a similar picture is found in the case of $\langle O_1^s \rangle$ and we thus refrain from showing the corresponding results.

Varying the input parameters within their intervals and combining all the corresponding uncertainties in quadrature, we obtain the following results for the matrix element $i \langle O^q_2 \rangle$, for both the modes considered, namely
\begin{align}
i \langle O_2^d \rangle & =  (0.24_{-0.22}^{+0.22}) \GeV^3\,, 
\qquad \qquad \bar B_s^0 \to D_s^+ \pi^- \,,
\label{eq:ME-O2d}
\\[2mm]
i \langle O_2^s \rangle & =  (0.24_{-0.19}^{+0.21}) \GeV^3\,,
\qquad \qquad \bar B^0 \to D^+ K^-\,,
\label{eq:ME-O2s}
\end{align}
where, as already discussed, the total uncertainties are strongly dominated by the limited accuracy of the parameter $\lambda_H^2$. 

We can now turn to discuss our results for the factorisable matrix element $\langle O_1^q \rangle$. The final sum rule is given in eq.~\eqref{eq:O1-SR} and includes the contribution of both the two- and three-particle LCDAs.  
The relative size of each of the LCDAs contributions is shown in the second column of Table~\ref{tab:part-contrib-LCDAs}. We find, as expected, that the dominant effect is due to $\phi_{\pm}$, with the twist-4 and twist-5 LCDAs $g_{\pm}$ yielding a smaller contribution. On the other hand, the three-particle LCDAs appear to be strongly suppressed, in consistency with what found e.g.\ in the LCSR study of the $B \to D$ form factors~\cite{Gubernari:2018wyi}.  As for the uncertainty budget, the LCSR prediction is extremely sensitive to the value of the non-perturbative parameter $\lambda_B$, while the dependence on the sum rule inputs i.e.\ the threshold continuum and the Borel parameters is found to be mild, and that on the parameters $\lambda_E^2$ and $\lambda_H^2$ very small. Furthermore, also in this case, the uncertainty due to $\mu$-variation is moderate. 

Varying all the input parameters within their intervals and again adding all individual uncertainties in quadrature, we obtain the following estimates 
of the factorisable matrix element $i \langle O_1^q \rangle$, for both the modes considered,~i.e.
\begin{eqnarray}
i \langle O_1^d \rangle & = & 
- (1.51^{+0.66}_{-0.61}) \GeV^3 \,,
\qquad \qquad \bar B_s^0 \to D_s^+ \pi^- \,,
\label{eq:ME-O1d}
\\[2mm] 
i \langle O_1^s \rangle & = & 
- (2.03^{+1.00}_{-0.75}) \GeV^3 \,,
\qquad \qquad \bar B^0 \to D^+ K^-\,.
\label{eq:ME-O1s}
\end{eqnarray}
Note that the above values are consistent with the QCDF results~\cite{Bordone:2020gao}, however the uncertainties are significantly larger. 

Before discussing our predictions for the branching fractions, two more remarks with respect to the error budget are in order. First, we emphasise that using other models for the LCDAs, like the local duality model, see Ref.~\cite{Braun:2017liq} for a detailed discussion, and Ref.~\cite{Lu:2018cfc} for new parametrisations of the twist-5 and twist-6 LCDAs, does not lead to any significant difference, within the quoted uncertainties, in the values for both the factorisable and non-factorisable matrix elements. Moreover, additional sources of uncertainties, like missing $1/m_b$ corrections to the expression in eq.~\eqref{eq:3-part-ME}, are also expected to be effectively covered by our large error ranges.

Combining the above results with the corresponding Wilson coefficients,
our estimates for the ratio of the non-factorisable over the factorisable parts of the amplitude for the $\bar B^0_s \to D_s^+ \pi^-$ and $\bar B^0 \to D^+ K^-$ decays read, respectively 
\begin{eqnarray}
&
\displaystyle \!\! \frac{C_2 \langle O_2^d \rangle}{C_1 \langle O_1^d \rangle} 
= 0.051^{+0.059}_{-0.052}\,,
\qquad \qquad \bar B_s^0 \to D_s^+ \pi^-\,,
&
\label{eq:R-O2-to-O1-d}
\\[2mm]
&
\displaystyle\frac{C_2 \langle O_2^s \rangle}{C_1 \langle O_1^s \rangle} =
0.039^{+0.042}_{-0.034}\,,
\qquad \qquad \bar B^0 \to D^+ K^-\,.
&
\label{eq:R-O2-to-O1-s}
\end{eqnarray}
The non-factorisable matrix element $\langle O_2^q\rangle$ is thus found to lead to a sizeable positive effect, of the order of few percent, to the total amplitude for both the non-leptonic decays considered. This is in perfect agreement with the first estimates of Ref.~\cite{Blok:1992na}, however in contrast with the results of Ref.~\cite{Bordone:2020gao}. On the other hand, the uncertainties in eqs.~\eqref{eq:R-O2-to-O1-d}, \eqref{eq:R-O2-to-O1-s} appear still very large, and are of the order of $100 \%$. It is worth pointing out that, despite computing the ratio of the two matrix elements within the same theoretical framework, we only obtain a minor reduction of the total uncertainty from the simultaneous variation of the common inputs. This follows from the large sensitivity of $\langle O_2^q \rangle$ and $\langle O_1^q \rangle$ on different non-pertubative parameters. That is, as already stressed, $\lambda_H^2$ for the former matrix element, and $\lambda_B$ for the latter. 
Furthermore, due to the stronger scale dependence of the Wilson coefficient $C_2$ compared to that of $C_1$, also the ratio $C_2/C_1$ does not provide a significant reduction of the total uncertainty. We note in particular that, because of the additional variation of the scale $\mu_b$, the relative uncertainty in the non-factorisable part of the amplitude becomes even larger.

The results in eqs.~\eqref{eq:ME-O2d}-\eqref{eq:ME-O1s} lead to the following predictions for the branching fractions
\begin{eqnarray}
{\rm Br} (\bar B^0_s \to D^+_s \, \pi^- \,) 
& = & 
(2.15^{+2.14}_{-1.35}) \times 10^{-3}, 
\label{eq:Bs-to-Ds-pi-LCSR} 
\\[2mm]
{\rm Br} (\bar B^0 \to D^+ K^-) 
& = & 
(2.04^{+2.39}_{-1.20}) \times 10^{-4},
\label{eq:B-to-D-K-LCSR}
\end{eqnarray}
in agreement with the experimental data shown in eqs.~\eqref{eq:Bs-to-Ds-pi-Exp}, \eqref{eq:B-to-D-K-Exp}, and also consistent with the QCDF results in eqs.~\eqref{eq:Bs-to-Ds-pi-QCDF}, \eqref{eq:B-to-D-K-QCDF}, although 
again within very large uncertainties. On the other hand, our central values are considerably lower than the latter.

Finally, note that naively combining our results in eqs.~\eqref{eq:R-O2-to-O1-d}, \eqref{eq:R-O2-to-O1-s}, with the QCDF prediction of the leading power amplitude for the corresponding decays, actually leads to a reduction of the observed tension with the data, despite the positive shift, due to the increased size of the uncertainties. However, we would like to emphasise that one should be careful when combining LCSR and QCDF results because of different assumptions adopted, e.g.\ the different treatment of the charm quark. 
\begin{table}[t]
\renewcommand{\arraystretch}{1.25}
    \centering
    \begin{tabular}{|c||c|c||c|}
    \hline 
     LCDA & 
     $\qquad i \langle O^d_1 \rangle \qquad$ &  
     $\qquad i \langle O^d_2 \rangle \qquad$ & 
     $i \langle O^d_2 \rangle $ (up to tw-4) \\
     \hline
     $\phi_+ $ & $-0.916 $  & -- & -- \\
     $\phi_- $ & $-0.863 $  & -- & -- \\
     $g_+ $    & $\phantom{-}0.243 $ & -- & -- \\ 
     $g_- $    & $\phantom{-}0.056 $ & -- & -- \\ 
     \hline 
     $\psi_V$ 
     & $-0.021$
     & $\phantom{-}0.264$ 
     & $\phantom{-}0.264$ \\
     $\psi_A$ 
     & $-0.006$
     & $\phantom{-}0.030$ 
     & $\phantom{-}0.030$ \\
     $\psi_{X_A}$ 
     & $-0.001$
     & $\phantom{-}0.000$ 
     & $\phantom{-}0.000$ \\
     $\psi_{Y_A}$ 
     & $\phantom{-}0.002$
     & $-0.016$ 
     & $-0.026$ \\
     $\psi_{\tilde X_A}$ 
     & $\phantom{-}0.000$
     & $-0.020$ 
     & $-0.020$\\
     $\psi_{\tilde Y_A}$ 
     & $-0.002$
     & $-0.018$ 
     & $-0.194$ \\
     $\psi_{W}$ 
     & $\phantom{-}0.001$
     & $\phantom{-}0.000$ 
     & $\phantom{-}0.020$ \\
     $\psi_{Z}$ 
     & $\phantom{-}0.000$
     & $\phantom{-}0.000$ 
     & $\phantom{-}0.007$ \\
     \hline 
     \hline
     $\Sigma$ 
     & $-1.507$  
     & $\phantom{-}0.240 $
     & $\phantom{-}0.081 $ \\
     \hline 
    \end{tabular}
    \caption{Partial contribution of the two- and three-particle LCDAs to the LCSR predictions for the matrix elements $\langle O_1^d \rangle$ and $\langle O_2^d \rangle$, in the case of the $\bar B_s^0 \to D_s^+ \pi^-$ decay. All results are in units of GeV$^{3}$ and correspond to using the exponential model for all the LCDAs, as well as central values for all inputs. The last column does not include the contribution of the twist-5 and twist-6 LCDAs in eqs.~\eqref{eq:phi-5-EM}-\eqref{eq:phi-6-EM}.} 
    \label{tab:part-contrib-LCDAs}
\end{table}
\section{Conclusion and outlook}
\label{sec:conclusion}
In this work we have presented new determinations, obtained within the framework of LCSR, of the non-factorisable contributions to the amplitude of the non-leptonic decays $\bar B_s^0 \to D_s^+ \pi^-$ and $\bar B^0 \to D^+ K^-$, due to soft gluon emission. The computation is based on the derivation of a LC-local OPE for a suitable three-point correlation function, and on the use of $B$-meson LCDAs. 
Our analysis, in particular, has raised several questions that have been overlooked in many previous similar studies, and that require further clarifications.
First, the fact that performing a double LC expansion of the correlation function seems to actually lead to a result which is not transversal. Second, in this case,
the dominant contribution to the correlator originates from 
generalised non-local three-particle $B$-meson matrix elements with non-aligned fields, which are still unknown in the literature for arbitrary Dirac structures. First steps in this direction have been taken in Ref.~\cite{Qin:2022rlk}. 
Third, we have found that truncating the expansion of the three-particle $B$-meson LCDAs at twist-4, i.e.\ neglecting the twist-5 and twist-6 DAs, seems to contradict the local limit of the corresponding non-local matrix element, and in addition lead to pronounced cancellations.
In our work, the first two points have been circumvented by employing a LC-local OPE, which, albeit less accurate, has allowed us to consistently compute the correlator in terms of known hadronic input functions. As for the third point, we have included the contribution of the twist-5 and twist-6 LCDAs in our analysis, thus ensuring the correct local limit for the three-particle matrix element, and also the lifting of the apparent cancellations. 
However, in light of the above findings, further investigations are certainly needed in order to improve the current understanding of these decays, as well as shed more light on the size of non-local hadronic effects in rare semileptonic $B$-meson decays. 

Another important result of the paper is the computation of the factorisable matrix elements for the decays $\bar B_s^0 \to D_s^+ \pi^-$ and $\bar B^0 \to D^+ K^-$, at LO-QCD accuracy, within LCSR, which represents the first determination using this framework. In this respect, it is important to stress that, despite so far the limited precision compared to QCDF at leading power, LCSR provides a well established method for the computation of the whole amplitude, including next-to-leading power effects, entirely within the same framework.

Our predictions, shown in eqs.~\eqref{eq:ME-O2d}-\eqref{eq:R-O2-to-O1-s}, indicate that the non-factorisable matrix element leads to a sizeable and positive contribution, of the order of few percent, to the amplitude for the decays $\bar B_s^0 \to D_s^+ \pi^-$ and $\bar B^0 \to D^+ K^-$, in consistency with the first estimates by Blok and Shifman~\cite{Blok:1992na}, but in contrast with the findings of Ref.~\cite{Bordone:2020gao}. On the other hand, we emphasise that the total uncertainties are also found to be very large, mainly due to the limited accuracy of many non-perturbative inputs, particularly those entering the parametrisation of the two- and three-particle $B$-meson LCDAs, i.e.\ $\lambda_B$, and $\lambda_H^2$. 

Finally, combining our results for the factorisable and non-factorisable matrix elements, we have also obtained new estimates for the branching fractions of the $\bar B_s^0 \to D_s^+ \pi^-$ and $\bar B^0 \to D^+ K^-$ decays, shown in eqs.~\eqref{eq:Bs-to-Ds-pi-LCSR}, \eqref{eq:B-to-D-K-LCSR}, respectively. 
Our predictions appear to be in good agreement with the corresponding experimental data, however, given the very large uncertainties, and the LO accuracy of the current analysis, we refrain from drawing any conclusion on the status of these observables, in light of the discrepancies found in Ref.~\cite{Bordone:2020gao}. We consider instead to be more justified to conclude with a comprehensive outlook for future studies and investigations. Specifically, in order to improve the present analysis, one would require:

\begin{itemize}
    \item[$\diamond$] 
    More accurate determination of the parameter $\lambda_B$, for both the $\bar B^0_d$ and $\bar B^0_s$ mesons, either by improving the current QCD sum rule analyses \cite{Braun:2003wx, Khodjamirian:2020hob} or by performing first Lattice QCD investigations. 
    Alternatively, stronger constraints on the size of these inputs could also be derived by extending the OPE-based studies of Refs.~\cite{Lee:2005gza, Kawamura:2008vq}, or from the anticipated data by the Belle~II collaboration on $B \to \gamma \ell  \nu$ decays~\cite{Beneke:2018wjp}.

    \item[$\diamond$] 
    Improved determination of the parameters $\lambda_E^2$ and $\lambda_H^2$ either within QCD sum rules or Lattice QCD, as well as the computation of the corresponding SU(3)$_F$-breaking effects which are, so far, still missing in the literature. 

    \item[$\diamond$] 
    Study of the generalised three-particle $B$-meson non-local matrix elements, with the light spectator quark and the gluon aligned on different light-cone directions. As already stressed in Ref.~\cite{Qin:2022rlk}, the knowledge of these novel soft functions would also be crucial in order to improve the current analyses of the non-local soft-gluon contributions in rare semileptonic $B$-meson decays, and thus to shed more light on the apparent tensions in $b \to s \ell^+ \ell^-$ transitions.

    \item[$\diamond$] 
    Further studies of higher-twist effects in the three-particle $B$-meson matrix elements, and of the corresponding LCDAs. The investigation of alternative models for the DAs would be important to reduce the corresponding model dependent uncertainty.

    \item[$\diamond$]
    Computation of NLO-QCD corrections in the OPE for both the factorisable and non-factorisable matrix elements, within LCSR.

    \item[$\diamond$]
    Alternative estimate of the factorisable and non-factorisable matrix elements using the LCSR framework with the light meson, i.e.\ the $\pi$- and $K$, LCDAs. This would in fact provide an important cross-check of our study, and allow one to circumvent the current challenges associated with the $B$-meson LCDAs.
    
\end{itemize}
\section*{Acknowledgements}
We are really grateful to Alexander Khodjamirian for  many insightful discussions and valuable comments. We also thank 
Nico Gubernari, Thorsten Feldmann, Alexander Lenz, and Danny van Dyk, for helpful discussions. In addition, we would like to thank Alexander Khodjamirian, and Alexander Lenz, for their constant support throughout this work and for carefully proofreading the manuscript.
We also acknowledge the TP1 group in Siegen for the useful feedback provided during the journal club, where this work has been informally presented. 
MLP wishes to thank Thorsten Feldmann, Peter Stangl, Javier Virto, and Roman Zwicky, for interesting discussions.
The work of MLP was funded by the Deutsche Forschungsgemeinschaft (DFG, German Research Foundation) - project number 500314741.
\appendix
\section{Conventions and definitions}
\label{app:A}
In this appendix, we collect the main conventions and definitions adopted throughout the paper. For the Levi-Civita tensor we use $\epsilon^{0123} = + 1$ which, together with $\gamma_5 = i \gamma^0 \gamma^1 \gamma^2 \gamma^3$, 
leads to 
\begin{equation}
{\rm Tr}[\gamma^\mu \gamma^\nu \gamma^\rho \gamma^\sigma \gamma_5] = - 4 i \epsilon^{\mu \nu \rho \sigma}\,.
\end{equation}
The SU(3)$_c$ generators in the fundamental representation $t^a_{ij}$ satisfy the completeness relation
\begin{equation}
    t^a_{ij}t^a_{lm} = \frac12\left(\delta_{im} \delta_{jl} - \frac{1}{N_c} \delta_{ij} \delta_{lm} \right)\,,
\end{equation}
and are normalised as Tr$[t^a t^b] = (1/2) \delta^{ab}$. The gluon field strength tensor is defined as $G_{\mu\nu} 
 = i [D_\mu, D_\nu]$, with the covariant derivative given by $D_\mu = \partial_\mu - i A_\mu(x)$. Note that the strong coupling $g_s$ is absorbed in the definition of the gluon field $A_\mu(x) = A_\mu^a(x) t^a$.
 
The matrix element of the axial and axial-vector currents $j_5^{D}(x) = i m_c \, \bar q \gamma_5 c$,  and $j_\mu^L(x) = \bar u \gamma_\mu \gamma_5 q$, with $q = \{d,s\}$, between the $D$- and the $L$-meson and the vacuum, are respectively defined as
\begin{equation}
 \langle 0 |j_5^D(x) | D (p) \rangle = m_D^2 f_D \, e^{- i p \cdot x}\,,
\qquad
 \langle 0 |j_\mu^L(x) | L (p) \rangle
 = 
i f_L \, p_\mu \, e^{- i p\cdot x}\,, 
\end{equation}
where $f_D$ and $f_L$ are the corresponding meson decay constants. Moreover, for the matrix element of the vector current $j_\mu(x) = \bar c \gamma_\mu b$ between a $B$- and a $D$-meson, we use the following parametrisation 
\begin{align}
\langle D (p) | j_\mu| B (p+q) \rangle 
&= 
f_+^{BD} (q^2) \left[2 p^\mu + \left(1 - \frac{m_B^2 - m_D^2}{q^2} \right) q^\mu \right] 
\nonumber \\[2mm]
&+  f_0^{BD} (q^2) \frac{m_B^2 - m_D^2}{q^2} q^\mu \,,
\end{align}
with $f_+^{BD}$ and $f_0^{BD}$ being, respectively, the vector and scalar form factors for the $B \to D$ transition. 

In order to write down the final sum rules, the following results for the imaginary part of the functions entering the OPE are used. From $\lim_{\varepsilon \to 0^+} 1/(x \pm i \varepsilon) = {\cal P}(1/x) \mp i \pi \delta(x)$, we obtain 
\begin{equation}
  {\rm Im} \frac{1}{\left(x - i \varepsilon \right)^n} = \pi \frac{(-1)^{n-1}}{(n-1)!} \, \delta^{(n-1)} \left( x \right)\,, \qquad n \geq 1\,,
   \label{eq:Im-delta}
\end{equation}
where $\delta^{(n-1)}(x)$ denotes the $(n-1)$-derivative of the delta function with respect to its argument. Furthermore, the analytic continuation of the logarithm function at negative values of the argument is defined as
\begin{equation}
    \ln(- x) =\ln|x| - i \pi \theta(x)\,,
    \label{eq:Im-log}
\end{equation}
so that the logarithms in eqs.~\eqref{eq:OPE-fac-2p}, \eqref{eq:OPE-fac-3p}, develop an imaginary part for $q^2>0$ equal to $- \pi$.
\section{B-meson LCDAs}
\label{app:B}
The non-local matrix element corresponding to both the light spectator quark and the gluon in the $B$-meson aligned along the same light-cone direction $n^\mu$ \footnote{The notation follows Ref.~\cite{Braun:2017liq} so, when comparing with section~\ref{sec:LC-dominance}, it is $n_+^\mu \equiv n^\mu$ and $n_-^\mu \equiv \bar n^\mu$.} in the rest frame of the heavy $B$-meson $v^\mu = (n^\mu + \bar n^\mu)/2 = (1,\vec 0 \,)$, can be parameterised in terms of eight three-particles LCDAs as~\cite{Braun:2017liq} \footnote{In the definition of the matrix element, the gauge link $[x,y] = P \exp \big\{ i \int_0^1 du \, (x-y)_\mu   A^\mu(u x + \bar u y)\big\}$, with $\bar u = 1 -u$, is always implicitly assumed. We note however that in the Fock-Schwinger gauge, which we use in our computation, this factor equals to unity.}
\begin{align}
& \langle 0 | \bar q_{\alpha} (n z_1) G_{\mu \nu} (n z_2) h_{v,\beta} (0) | \bar B (v) \rangle   = 
  \frac12 F_B(\mu) 
 \Big\{ P_+ \big[(v_\mu \gamma_\nu - v_\nu \gamma_\mu)(\Psi_A - \Psi_V) - i \sigma_{\mu \nu} \Psi_V
    \nonumber \\[1mm]
    & - (n_\mu v_\nu - n_\nu v_\mu) \Psi_{X_A} + (n_\mu \gamma_\nu - n_\nu \gamma_\mu) (\Psi_{W} + \Psi_{Y_A}) + i \epsilon_{\mu \nu \eta \tau} n^\eta v^\tau \gamma_5 
    \Psi_{\tilde{X}_A} - i \epsilon_{\mu \nu \eta \tau} n^\eta \gamma^\tau 
    \nonumber \\[1mm]
    & 
    \times \gamma_5  \Psi_{\tilde{Y}_A} - (n_\mu v_\nu - n_\nu v_\mu) \slashed n \Psi_W + (n_\mu \gamma_\nu - n_\nu \gamma_\mu) \slashed n \Psi_Z
    \big] \gamma_5 \Big\}_{\beta \alpha}\!\!(z_1, z_2; \mu) \,,
    \label{eq:app:3-Part-ME}
\end{align}
where $\alpha, \beta,$ denote spinor indices, $h_v(x) = e^{i m_b v \cdot x} b(x) + {\cal O}(1/m_b)$ is the HQET field, see e.g.\ the review~\cite{Neubert:1993mb}, and the sign difference with respect to Ref.~\cite{Braun:2017liq} in the coefficients of the Levi-Civita tensor follows from using the opposite convention for $\epsilon_{\mu \nu \rho \sigma}$, cf.~appendix~\ref{app:A}. Moreover, we point out the change of notation for some of the DAs as compared to Ref.~\cite{Braun:2017liq}, i.e.\ $ X_A \to \Psi_{X_A}$ etc.\

In eq.~\eqref{eq:app:3-Part-ME}, the two parameters $z_1, z_2,$ specify, respectively, the position of the light quark and of the gluon field on the light-cone vector $n^\mu$. Performing a Fourier transform, each LCDA can be expressed in terms of the corresponding momentum space distributions as
\begin{equation}
\Psi(z_1, z_2) = \int_0^\infty \!\!\! d \omega_1 \int_0^\infty \!\!\!  d \omega_2 \,\,  e^{-i \omega_1 z_1 -i \omega_2 z_2} \,\, \psi(\omega_1, \omega_2)\,,
\label{eq:Fourier-transf}
\end{equation}
$\Psi = \{\Psi_V, \Psi_A, \ldots\}$. Note that following Ref.~\cite{Braun:2017liq}, we also adopt the convention that DAs in coordinate space are written in upper case, whereas the lower case is used for the momentum space representations.
 
In order to reorganise eq.~\eqref{eq:app:3-Part-ME} in terms of its twist expansion rather than its Lorentz decomposition, in Ref.~\cite{Braun:2017liq}, the DAs appearing in eq.~\eqref{eq:app:3-Part-ME} have been recast in terms of LCDAs of definite collinear twist. Specifically~\cite{Braun:2017liq}
\begin{align}
    &\Psi_A(z_1,z_2) = \frac{\Phi_3 + \Phi_4}{2}\,,
    \nonumber \\[1mm]
    &\Psi_V(z_1,z_2) = \frac{\Phi_4 - \Phi_3}{2}\,,
    \nonumber \\[1mm]
    &\Psi_{X_A} (z_1,z_2) = \frac{-\Phi_3 - \Phi_4 + 2 \Psi_4}{2}\,,
    \nonumber \\[1mm]
    &\Psi_{\tilde X_A}(z_1,z_2) = \frac{-\Phi_3 + \Phi_4 - 2 \tilde \Psi_4}{2}\,,
    \label{eq:LCDAs-twist}
    \\[1mm]
    &\Psi_{Y_A} (z_1,z_2) = \frac{-\Phi_3 - \Phi_4 + \Psi_4- \Psi_5}{2}\,,
    \nonumber \\[1mm]
    &\Psi_{\tilde Y_A}(z_1,z_2) = \frac{-\Phi_3 + \Phi_4 - \tilde \Psi_4 + \tilde \Psi_5}{2}\,,
    \nonumber \\[1mm]
    &\Psi_{W} (z_1,z_2) = \frac{\Phi_4 - \Psi_4 - \tilde \Psi_4 + \tilde \Phi_5 +  \Psi_5 + \tilde \Psi_5}{2}\,,
    \nonumber \\[1mm]
    &\Psi_{Z} (z_1,z_2) = \frac{-\Phi_3 + \Phi_4- 2 \tilde \Psi_4 + \tilde \Phi_5 + 2 \tilde \Psi_5 - \Phi_6}{4}\,,
    \nonumber
\end{align}
where $\Phi_3$, and $\Phi_4, \Psi_4, \tilde \Psi_4$, are LCDAs of twist-3 and twist-4, whereas $\tilde \Phi_5, \Psi_5, \tilde \Psi_5$, and $\Phi_6$, are of twist-5 and twist-6, respectively. 
 
At leading order in HQET, the three-particle matrix elements required for the computation of both the factorisable and non-factorisable amplitudes can be then simply derived from eqs.~\eqref{eq:app:3-Part-ME}. In doing so, one obtains 
additional factors of the type $(v \cdot x)^{-1}$ and $(v \cdot x)^{-2}$, which can be simplified by introducing respectively the replacements
 \begin{equation}
     \psi(\omega_1, \omega_2) \to i (v \cdot x) \bar \psi(\omega_1, \omega_2)\,, 
     \qquad
     \psi(\omega_1, \omega_2) \to - (v \cdot x)^2 \, \bar{\bar{\psi}}(\omega_1, \omega_2)\,,
\label{eq:Psi-bar-sub}     
 \end{equation}
with 
\begin{equation}
    \bar{\psi}(\omega_1, \omega_2) \equiv 
    \int_0^{\omega_1} \!\! d \eta \, 
      \psi(\eta, \omega_2) \,,
\qquad
    \bar{\bar{\psi}}(\omega_1, \omega_2) \equiv 
    \int_0^{\omega_1} \!\! d \eta
    \int_0^{\eta} \!\! d \eta^\prime \, 
      \psi(\eta^\prime, \omega_2) \,.
\label{eq:Psi-bar-def}      
\end{equation}
The results in eq.~\eqref{eq:Psi-bar-sub}, \eqref{eq:Psi-bar-def}, can be easily derived by using the identities
\begin{equation}
   \int d^4 x \, f(x) \int_0^\infty \!\! d \omega_1 \, \frac{d}{d \omega_1}  \Big[ e^{- i \omega_1 v \cdot x} \int_0^{\omega_1} \!\! d \eta \,\,  \psi(\eta, \omega_2)
    \Big] =0 \,,
    \label{eq:psi-bar-proof}
\end{equation}
\begin{equation}
   \int d^4 x \, f(x) \int_0^\infty \!\! d \omega_1 \, \frac{d^2}{d \omega_1^2}  \Big[ e^{- i \omega_1 v \cdot x} \int_0^{\omega_1} \!\! d \eta  \int_0^{\eta} \!\! d \eta^\prime \,\,  \psi(\eta^\prime, \omega_2)
    \Big] =0 \,,
    \label{eq:psi-bar-bar-proof}
\end{equation}
where $f(x)$
absorbs the remaining $x$-dependence in the correlation function.
Note that eqs.~\eqref{eq:psi-bar-proof}, \eqref{eq:psi-bar-bar-proof}, follow from the fact that
the boundary terms are always zero. Specifically, they vanish when $\omega_1 \to 0$ due to the integration over $\eta$, and also when $\omega_1 \to \infty$, because of the exponential suppression of the integral over $x^\mu$, in accordance with the Riemann–Lebesgue theorem. 

In our computation of the non-factorisable amplitude, we actually need the corresponding three-particle matrix element with the gluon field fixed at the origin, cf.~eq.~\eqref{eq:F-tilde-1}. Setting $z_2 = 0$ in eq.~\eqref{eq:app:3-Part-ME}, the integral over $\omega_2$ in eq.~\eqref{eq:Fourier-transf} can be readily performed. To this end, we introduce the compact notation
\begin{equation}
    \hat \psi(\omega_1) \equiv \int_0^\infty d \omega_2 \,\, \psi(\omega_1,\omega_2)\,.
    \label{eq:psi-hat}
\end{equation}

Finally, a new parametrisation of the two-particle $B$-meson matrix element including higher-twist DAs, at leading order in HQET, has been obtained in Ref.~\cite{Braun:2017liq}
\begin{align}
 \langle 0| \bar q_\alpha (x) h_{v, \beta}(0)|\bar B(v) \rangle & = - \frac{i}{2} F_{B}(\mu) \sqrt{m_{B}}  \int_0^\infty \!\! d \omega  \,  e^{- i \omega v \cdot x} \Big\{ \big(\phi_+ + x^2 g_+) P_+ \gamma_5 
\nonumber \\[1mm]
& - \frac{1}{2 (v \cdot x)} \big[(\phi_+ - \phi_-) + x^2 ( g_+ -  g_-)] P_+ \, \slashed x \, \gamma_5 \Big\}_{\beta \alpha}\!\!(\omega ; \mu)\,,
\label{eq:app-2-part-ME-fac}
\end{align}
where $\phi_+$, $\phi_-$, are LCDAs of twist-2 and twist-3, whereas $g_+$, $g_-$, are of twist-4 and twist-5, respectively. The expression in eq.~\eqref{eq:2-part-ME-fac} immediately follows from eq.~\eqref{eq:app-2-part-ME-fac}, taking into account that
\begin{equation}
\phi(\omega) \to i (v \cdot x) \bar \phi(\omega)\,, \qquad {\rm with}\qquad \bar \phi (\omega) \equiv \int_0^\omega d \eta\, \phi(\eta)\,,
\label{eq:phi-bar-app}
\end{equation}
with $\phi= \{\phi_+, g_+, \ldots \}$.
\section{Computation of the loop integral in coordinate space}
\label{app:loop-func-coord-space}
The one-loop integral in eq.~\eqref{eq:Correlator-O1T-2}, can be also explicitly computed in coordinate space, since in the limit of massless quark, the expression of the corresponding propagator simplifies and does not contain Bessel functions. In fact, in dimensional regularisation, with $d = 4 - 2 \epsilon$, the local expansion of a massless quark propagator, up to leading one-gluon contributions, reads~\cite{Balitsky:1987bk} 
\begin{align}
S^{(q)}_{ij} (x,y) &=  \frac{\Gamma (d/2)}{2 \pi^2} \frac{\slashed x - \slashed y}{\big[ -(x-y)^2\big]^{d/2}} \, \delta_{ij} 
\nonumber \\[2mm]
& + \frac{\Gamma (d/2-1)}{32 \pi^2} \frac{(\slashed x - \slashed y) \, \sigma^{\mu \nu} + \sigma^{\mu \nu} (\slashed x -\slashed y)}{\big[ -(x-y)^2\big]^{d/2-1}} G_{\mu \nu}^a \, t_{ij}^a + \ldots \,,
\label{eq:massless-prop-d-dim-coord}
\end{align}
where $q =\{u, d, s\}$, $\Gamma(z)$ is the gamma function, and the ellipses denote higher order corrections with at least one covariant derivative of the gluon field strength tensor. 
Substituting eq.~\eqref{eq:massless-prop-d-dim-coord} into eq.~\eqref{eq:Correlator-O1T-2}, the integration over $y^\mu$ can be easily performed by taking into account the following results~\cite{Novikov:1983gd}
\begin{align}
\int d^d y \, e^{i q \cdot y} \frac{1}{(-y^2)^a}
& = -i \, 2^{d-2a} \pi^{d/2} \, \frac{\Gamma(d/2 - a)}{\Gamma(a)} \, (-q^2)^{a- d/2}\,,
\label{eq:dDy-scalar} 
\\[2mm]
\int d^d y \, e^{i q \cdot y} \frac{y^\mu y^\nu}{(-y^2)^a} 
& = 
i \, 2^{d-2 a + 1} \pi^{d/2}\, \frac{\Gamma(d/2 - a)}{\Gamma(a)} \, \left(a - \frac{d}{2} \right) (-q^2)^{a - d/2 - 2} 
\nonumber \\
& \times  \left[ \left(a - \frac{d}{2} - 1 \right) 2 q^\mu q^\nu + q^2 g^{\mu \nu} \right]\,,
\label{eq:dDy-tensor}
\end{align}
where $a = d - 1$, and eq.~\eqref{eq:dDy-tensor} has been obtained by differentiating twice both sides of eq.~\eqref{eq:dDy-scalar} with respect to $q^\mu$.
Performing the calculation in NDR, the divergent $1/\epsilon$ contributions cancel when considering the gluon emission from both the light-quark propagators, leading to the finite result shown in eq.~\eqref{eq:I-alphabetamu}. Finally, the computation of the loop integral in eq.~\eqref{eq:Correlator-O1-2} proceeds in an analogous way, although now only the first line of eq.~\eqref{eq:massless-prop-d-dim-coord} contributes.
\section{Inverse Fourier transform of Bessel functions}
\label{app:D}
In this appendix, we list the results for the tensor integrals introduced in eqs.~\eqref{eq:int-x-K1}, \eqref{eq:int-x-K2}, \eqref{eq:int-x-K0}. Starting with those of lowest rank
\begin{align}
& \int d^4 x \,\, e^{i p \cdot x}\, K_0 (m \sqrt{-x^2}) = - 8 \pi^2 i \, \frac{1}{(p^2 - m^2 + i \varepsilon)^2}\,,
\label{eq:int-K0-app}
\\[2mm]
& \int d^4 x \,\, e^{i p \cdot x}\,\, \frac{K_1(m \sqrt{-x^2})}{\sqrt{-x^2}} = \frac{4 \pi^2 i}{m} \frac{1}{ p^2 - m^2 + i \varepsilon}\,,
\label{eq:int-K1-app}
\\[2mm]
& \int d^4 x \,\, e^{i p \cdot x}\,\, \frac{K_2(m \sqrt{-x^2})}{x^2} x^\mu = - \frac{4  \pi^2}{m^2} \frac{ p^\mu}{ p^2 - m^2 + i \varepsilon}\,,  
\label{eq:int-K2-x-app}
\end{align}
the remaining tensor integrals can be obtained by  differentiating multiple time eqs.~\eqref{eq:int-K0-app}-\eqref{eq:int-K2-x-app} with respect to the four-momentum $p^\mu$. This gives
\begin{align}
& \int d^4 x \,\, e^{i p \cdot x}\,\, K_0 (m \sqrt{-x^2}) \, x^\mu
= 32 \pi^2 \frac{p^\mu}{(p^2 - m^2 + i \varepsilon )^3}\,,
\label{eq:int-K1-x-app}
\\[3mm]
& \int d^4 x \,\, e^{i p \cdot x}\,\, K_0 (m \sqrt{-x^2}) \, x^\mu x^\nu
= 32 \pi^2 i \, \left[\frac{6 p^\mu p^\nu - (p^2 - m^2) g^{\mu \nu}}{(p^2 - m^2 + i \varepsilon )^4}\right]\,,
\label{eq:int-K1-x-app}
\\[3mm]
& \int d^4 x \,\, e^{i p \cdot x}\,\, \frac{K_1(m \sqrt{-x^2})}{\sqrt{-x^2}} x^\mu
= - \frac{8 \pi^2}{m} \frac{p^\mu}{(p^2 - m^2 + i \varepsilon )^2}\,,
\label{eq:int-K1-x-app}
\\[3mm]
& \int d^4 x \,\, e^{i p \cdot x}\,\, \frac{K_1(m \sqrt{-x^2})}{\sqrt{-x^2}} x^\mu x^\nu
= \frac{8 \pi^2 i}{m} \left[ \frac{( p^2 - m^2) g^{\mu \nu} - 4 p^\mu  p^\nu }{( p^2 - m^2 + i \varepsilon)^3} \right]\,,
\label{eq:int-K1-x2-app}
\\[3mm]
& \int d^4 x \,\, e^{i p \cdot x}\,\, \frac{K_1(m \sqrt{-x^2})}{\sqrt{-x^2}} x^2 x^\mu 
=  192 \pi^2 m  \frac{p^\mu}{(p^2 - m^2 + i \varepsilon)^4} \,,
\label{eq:int-K1-x3-app}
\\[3mm]
&\int d^4 x \,\, e^{i p \cdot x}\,\, \frac{K_2(m \sqrt{-x^2})}{x^2} x^\mu x^\nu = \frac{4  \pi^2 i}{m^2} \left[ \frac{( p^2 - m^2) g^{\mu \nu} - 2  p^\mu p^\nu }{( p^2 - m^2 + i \varepsilon)^2} \right]\,,  
\label{eq:int-K2-x2-app}
\\[3mm]
&\int d^4 x \,\, e^{i p \cdot x}\,\, \frac{K_2(m \sqrt{-x^2})}{x^2} x^\mu x^\nu x^\rho = - \frac{8 \pi^2}{m^2} \left[ \frac{(p^2 - m^2) g^{\{ \mu\nu} p^{\rho \}} - 4 p^\mu p^\nu p^\rho }{(p^2 - m^2 + i \varepsilon)^3} \right]\,,
\label{eq:int-K2-x3-app}
\\[3mm]
&\int d^4 x \,\, e^{i p \cdot x}\,\, K_2(m \sqrt{-x^2}) x^\mu x^\nu  = - \frac{16 \pi^2 i}{m^2} \left[ \frac{(p^2 - 4 m^2) (4 p^\mu p^\nu - g^{\mu \nu} p^2) - 3 m^4 g^{\mu \nu}}{(p^2 -  m^2 + i \varepsilon)^4} \right]\,,
\label{eq:int-K2-x4-app}
\end{align}
where the curly brackets in eq.~\eqref{eq:int-K2-x3-app} denote the symmetrisation of the tensor $g^{\mu \nu} p^\rho $ with respect to the three Lorentz indices.
\section{Results for the OPE coefficients}
\label{app:E}
The $c_n^{\hat \psi} (\omega_1, q^2)$ in eq.~\eqref{eq:Fq-OPE} read respectively \footnote{Here and in the rest of the section, we only show the non-vanishing coefficients.}
\begin{align}
c_1^{\hat \psi_V}(\omega_1, q^2) 
& = - \frac{1}{16 \pi ^2 \left(m_B - \omega_1\right)^3}
\Bigl(3 m_B^2 + 4 m_B m_c - 6 m_B \, \omega_1 + m_c^2 - q^2  
\nonumber \\[1mm]
&  - 4 m_c \, \omega_1 + 3 \, \omega_1^2 \Bigr) \Bigl(m_B^3 - 2 m_B^2 \, \omega_1 - m_B (m_c^2 + q^2 - \omega_1^2) + 2 q^2 \omega_1 \Bigr)\,,
\\[4mm]   
c_1^{\hat \psi_A}(\omega_1, q^2) 
& = - \frac{1}{16 \pi ^2 \left(m_B -\omega_1\right)^3} 
\Bigl(m_B^5 - 4 m_B^4 \, \omega_1 - 2 m_B^3 \left(m_c^2 - 3 \,\omega_1^2\right)
\nonumber \\[1mm]
& + 2 m_B^2 \, \omega_1 \left(2 m_c^2 + q^2 - 2 \, \omega_1^2\right) 
+ m_B (m_c^4 - 2 m_c^2 \, \omega_1^2  - q^4 - 4 q^2 \omega_1^2 + \omega_1^4)
\nonumber \\[1mm]
& + 2 q^2 \omega_1 (-m_c^2 + q^2 +\omega_1^2) \Bigr)\,,
\\[4mm]
c_1^{\bar{\hat \psi}_{Y_A}}(\omega_1, q^2)
& = \frac{1}{4 \pi ^2 \left(m_B-\omega_1\right)^2}
   \Bigl(m_B^3 - 2 m_B^2\,\omega_1 - m_B (m_c^2+q^2-\omega_1^2) +3 q^2 \omega_1 \Bigr)\,, 
\\[4mm]
c_2^{\bar{\hat \psi}_{Y_A}}(\omega_1, q^2) 
& = - \frac{m_B}{8 \pi ^2 \left(m_B-\omega_1\right)^3}
\Bigl(m_B^2 - 2 m_B \, \omega_1 - m_c^2 - q^2 + \omega_1^2 \Bigr) 
\nonumber \\[1mm]
& \times \Bigl(m_B^3 - 2 m_B^2 \, \omega_1 - m_B (m_c^2 + q^2 - \omega_1^2) + 2 q^2 \omega_1 \Bigr)\,,
\\[4mm]
c^{\bar{\hat \psi}_{\tilde X_A}}_1(\omega_1, q^2) 
&  = - \frac{m_B + m_c - \omega_1}{4 \pi ^2 \left(m_B-\omega_1\right)^3}
\bigl(m_B^2 m_c - m_B m_c (m_c + \omega_1) + q^2 \omega_1 \bigr)\,,
\\[4mm]
c^{\bar{\hat \psi}_{\tilde X_A}}_2(\omega_1, q^2) 
& = \frac{m_B (m_B+m_c-\omega_1)}{8 \pi^2 \left(m_B-\omega _1\right)^4}
 \Bigl(m_B^2  - 2 m_B (m_c + \omega_1) + m_c^2 - q^2
\nonumber \\[1mm]
&  + 2 m_c \, \omega_1 + \omega_1^2\Bigr) 
\Bigl(m_B^3 - 2 m_B^2 \, \omega_1 - m_B (m_c^2 + q^2 - \omega_1^2) + 2 q^2 \omega_1\Bigr)\,,   
\\[4mm]
c^{\bar{\hat \psi}_{\tilde Y_A}}_1(\omega_1, q^2) 
& =  \frac{1}{4 \pi ^2 \left(m_B-\omega _1\right)^2}
\Bigl(m_B^3 + 2 m_B^2 (m_c-\omega_1)  
\nonumber \\[1mm]
& + m_B (m_c^2 - 2 m_c\,\omega_1 - q^2 + \omega_1^2) + q^2 \omega_1 \Bigr)\,,
\\[4mm]
c^{\bar{\hat \psi}_{\tilde Y_A}}_2(\omega_1, q^2) 
& = \frac{m_B}{8 \pi ^2 \left(m_B-\omega _1\right)^3}
\Bigl(m_B^2 + 4 m_B m_c - 2 m_B \, \omega_1 + 3 m_c^2 - q^2
\nonumber \\[1mm]
& - 4 m_c \, \omega_1 + \omega_1^2 \Bigr)
\Bigl(m_B^3 - 2 m_B^2 \, \omega_1 - m_B (m_c^2 + q^2 - \omega_1^2) + 2 q^2 \omega_1 \Bigr)\,, 
\end{align}
\begin{align}
   c_1^{\bar{\hat\psi}_{W}}(\omega_1, q^2) 
   & = \frac{1}{4 \pi ^2 \left(m_B - \omega_1\right)^2}
   \Bigl(m_B^3 - 2 m_B^2 \, \omega_1 - m_B (m_c^2 + q^2 -\omega_1^2)
   + 3 q^2 \omega_1 \Bigr)\,, 
   \\[4mm]
   c_2^{\bar{\hat\psi}_{W}}(\omega_1, q^2) 
   & = - \frac{m_B}{8 \pi ^2 \left(m_B-\omega_1\right)^3}
   \Bigl(m_B^2 - 2 m_B \, \omega_1 - m_c^2 - q^2 + \omega_1^2 \Bigr)
   \nonumber \\[1mm]
   & \times \Bigl(m_B^3 - 2 m_B^2 \, \omega_1 - m_B (m_c^2 + q^2 - \omega_1^2) + 2 q^2 \omega_1 \Bigr)\,,
   \\[4mm]
c^{\bar{\bar{\hat \psi}}_{Z}}_1(\omega_1, q^2) 
& =  \frac{m_B m_c}{2 \pi ^2  \left(m_B-\omega_1\right)^2} \,,  
\\[4mm]
c^{\bar{\bar{\hat \psi}}_{Z}}_2(\omega_1, q^2) 
& =\frac{m_B m_c \, q^2 \omega_1}{\pi ^2 \left(m_B - \omega_1\right)^3}\,,
\\[4mm]
c^{\bar{\bar{\hat \psi}}_{Z}}_3(\omega_1, q^2) 
& = - \frac{m_B^2 m_c}{2 \pi ^2 \left(m_B-\omega_1\right)^4}
   \Bigl(m_B^2  - 2 m_B \, \omega_1 - m_c^2 - q^2 + \omega_1^2 \Bigr)
   \nonumber \\[1mm]
& \times \Bigl(m_B^3 - 2 m_B^2 \, \omega_1 - m_B (m_c^2 + q^2 - \omega_1^2) + 2 q^2 \omega_1 \Bigr)\,.
\end{align}

\begin{center}
$***$
\end{center}
The coefficients $c_n^\phi(\omega, q^2)$ in eq.~\eqref{eq:OPE-fac-2p} read respectively
\begin{align}
    c_1^{\phi_+}(\omega, q^2) 
    & = \frac{m_B+m_c-\omega}{8 \pi ^2 \left(m_B-\omega \right)^2}
    \Bigl(m_B^3 -2 \omega  m_B^2 - m_B (m_c^2+q^2-\omega ^2)+2q^2 \omega \Bigr)\,,
\\[4mm]
   c_1^{\bar \phi_+}(\omega, q^2) 
   & = - \frac{1}{8 \pi ^2 \left(m_B-\omega \right)^2}
   \Bigl(m_B^3 + m_B^2 (m_c -2 \omega ) 
- m_B \left(\omega (m_c -\omega) + q^2 \right) + 2 q^2 \omega \Bigr)\,,
\\[4mm]
 c_2^{\bar \phi_+}(\omega, q^2) &
 = - \frac{m_B m_c (m_B+m_c-\omega)}{8 \pi ^2 \left(m_B-\omega\right)^3}
  \Bigl(m_B^3 -2 \omega 
   m_B^2 - m_B (m_c^2+q^2-\omega ^2) + 2 q^2 \omega \Bigr)\,
 \\[4mm]
c_1^{\bar \phi_-}(\omega, q^2) &=\frac{1}{8 \pi ^2 \left(m_B-\omega \right)^2}
  \Bigl(m_B^3 + m_B^2 (m_c-2 \omega) - m_B \left(\omega  (m_c-\omega) + q^2 \right)+2 q^2 \omega \Bigr)\,,
\\[4mm]
 c_2^{\bar \phi_-}(\omega, q^2) &= \frac{m_B m_c \left(m_B+m_c-\omega \right)}{8 \pi ^2 \left(m_B-\omega \right)^3}
   \Bigl(m_B^3 - 2 \omega m_B^2 -m_B (m_c^2+q^2-\omega ^2)+2 q^2 \omega \Bigr)\,,
\\[4mm]
   c_1^{g_+}(\omega, q^2) 
   & = -\frac{m_B}{2 \pi ^2 \left(m_B-\omega \right)}\,,
\end{align}
\begin{align}
   c_2^{g_+}(\omega, q^2) 
   & = -\frac{m_B}{2 \pi ^2 \left(m_B-\omega \right)^3}
   \Bigl(m_B^4 - 3 \omega  m_B^3 + m_B^2 \left(m_c^2-q^2+3\omega ^2\right) 
 \nonumber \\[1mm]
   & + m_B \left(-\omega  m_c^2+2 m_c^3+3 q^2 \omega -\omega ^3\right) -2 q^2 \omega ^2 \Bigr)\,,
   \\[4mm]
   c_3^{g_+}(\omega, q^2) 
   & = -\frac{m_B^2 m_c^2 (m_B + m_c - \omega)}{\pi ^2 \left(m_B-\omega \right)^4}
   \Bigl(m_B^3 - 2 \omega m_B^2 - m_B (m_c^2 + q^2 - \omega^2) + 2 q^2 \omega \Bigr)\,,
\\[4mm]
   c_3^{\bar g_+}(\omega, q^2) 
   & =\frac{3 m_B^3 m_c^3 \left(m_B+m_c-\omega \right)}{\pi ^2 \left(m_B-\omega \right)^4}\,,
\\[4mm]
   c_4^{\bar g_+}(\omega, q^2) 
   & =\frac{3 m_B^3 m_c^3 \left(m_B+m_c-\omega \right)}{\pi ^2 \left(m_B-\omega \right)^5} \Bigl(m_B^3 -2 \omega m_B^2 -m_B \left(m_c^2+q^2-\omega ^2\right)+2 q^2 \omega \Bigr)\,,
\\[4mm]
    c_3^{\bar g_-}(\omega, q^2) 
    & = -\frac{3 m_B^3 m_c^3 \left(m_B+m_c-\omega \right)}{\pi ^2 \left(m_B-\omega \right)^4}\,,
\\[4mm]
   c_4^{\bar g_-}(\omega, q^2) 
   & = -\frac{3 m_B^3 m_c^3 \left(m_B+m_c-\omega \right)}{\pi ^2 \left(m_B-\omega \right)^5} \Bigl(m_B^3 -2 \omega m_B^2 - m_B \left(m_c^2+q^2-\omega ^2\right) + 2 q^2 \omega \Bigr)\,.
\end{align}

\begin{center}
$***$
\end{center}
Finally, the coefficients $c_n^\psi(u, \omega, q^2)$ in eq.~\eqref{eq:OPE-fac-3p} read respectively
\begin{align}
   c_1^{\psi_V} (u, \omega, q^2)  
   & = \frac{1}{8 \pi ^2 \left(m_B-\omega \right)^3}
  \Bigl(m_B \left(m_B+m_c-\omega \right) \bigl(m_B (4 u-1) + 4 m_c (1-u)
\nonumber  \\[1mm]
  & -4 u \omega +\omega \bigr)+4 q^2 (u-1) \omega \Bigr)\,,
\\[4mm]
   c_2^{\psi_V} (u, \omega, q^2)  
   & =\frac{m_B}{8 \pi ^2 \left(m_B-\omega \right)^4}
  \Bigl(m_B \left(-2 \omega  m_B+m_B^2-m_c^2+\omega ^2\right)-q^2 \left(m_B-2 \omega \right) \Bigr)  
\nonumber \\[1mm]
   & \times \Bigl(\left(m_B+m_c-\omega \right) \left(m_B (2 u - 1) + 2 m_c (1-u) - (2 u + 1) \omega \right) 
\nonumber \\
& \quad + 2 q^2 (u-1)\Bigr)\,,
\\[4mm]
c_1^{\psi_A} (u, \omega, q^2) 
    & = \frac{1}{8 \pi^2 \left(m_B - \omega \right)^3}
    \Bigl(m_B \bigl(3 m_B m_c - 2 m_B (2 u \omega + \omega) + (2 u + 1) m_B^2
\nonumber \\[1mm]
    & + 4 (u - 1) m_c^2 - 3 \omega m_c + (2 u + 1) \omega^2 \bigr) - 4 q^2 (u-1) \omega \Bigr)\,,
\end{align}
\begin{align}
   c_2^{\psi_A} (u, \omega, q^2)  
   & = \frac{m_B}{8 \pi ^2 \left(m_B-\omega \right)^4}
   \Bigl(m_B \left(-2 \omega  m_B+m_B^2-m_c^2+\omega ^2\right)-q^2 \left(m_B-2 \omega \right)\Bigr)
\nonumber \\[1mm]
   &\times \Bigl(m_B \left(3 m_c-8 u \omega +2 \omega \right)+(4 u-1) m_B^2
   + 2 m_c^2 (u - 1) - 3 \omega m_c
\nonumber \\[1mm]
   & -2 q^2 (u - 1) + 4 u \omega^2 - \omega^2 \Bigr)\,,
   \\[4mm]
   c_1^{\bar \psi_{X_A}} (u, \omega, q^2)  
   & = - \frac{m_B}{8 \pi^2 \left(m_B - \omega \right)^3}
   \Bigl((2 u-1) m_B + 2 m_c-2 u \omega +\omega \Bigr)\,,
\\[4mm]
   c_2^{\bar \psi_{X_A}} (u, \omega, q^2)  
   & = \frac{m_B}{8 \pi ^2 \left(m_B-\omega \right)^4}
   \Bigl(q^2 \bigl(m_B (\omega -2 u \omega) + (1-2 u) m_B^2
    \nonumber  \\[1mm]
   & + 2 \omega  \bigl((2 u-1) \omega - 2 m_c \bigr)\bigr) 
   + m_B \left(-2 \omega  m_B+m_B^2-m_c^2+\omega ^2\right) 
\nonumber \\[1mm]
   & \quad \times\bigl((2 u-1) m_B-4 m_c -2 u \omega +\omega \bigr)\Bigr)\,,
\\[4mm]   
   c_3^{\bar \psi_{X_A}} (u, \omega, q^2)  
   & = \frac{m_B^2 \left(m_B \left(-2 \omega  m_B+m_B^2-m_c^2+\omega ^2\right)-q^2 \left(m_B-2 \omega
   \right)\right)}{4 \pi ^2 \left(m_B-\omega
   \right)^5} 
\nonumber  \\[1mm]
   & \times 
   \Bigl(\left(2 \omega  m_B-m_B^2+m_c^2-\omega ^2\right) \left(-2 u m_B+m_B+m_c+(2 u-1) \omega \right)
\nonumber   \\[1mm]
   & -q^2 \bigl((2 u-1) m_B+m_c-2 u \omega +\omega \bigr)\Bigr)\,,
\\[4mm]
   c_2^{\bar \psi_{Y_A}} (u, \omega, q^2)  
   & = \frac{3 m_B^2 m_c}{2 \pi ^2 \left(m_B-\omega \right)^3}
   \Bigl(m_B+(2 u-1) m_c-\omega \Bigr)\,,
\\[4mm]
   c_3^{\bar \psi_{Y_A}} (u, \omega, q^2)  
   & = \frac{3 m_B^2 m_c \left(m_B \left(-2 \omega  m_B+m_B^2-m_c^2+\omega ^2\right)-q^2 \left(m_B-2 \omega
   \right)\right)}{2 \pi ^2 \left(m_B-\omega \right)^4} 
\nonumber \\[1mm]
   & \times \Bigl(m_B+(2 u-1) m_c-\omega \Bigr)\,,
 \\[2mm]
   c_1^{\bar \psi_{\tilde X_A}} (u, \omega, q^2)  
   & =\frac{m_B \left(m_B+2 m_c-\omega \right)}{8 \pi ^2 \left(m_B-\omega \right)^3}\,,
\\[4mm]
   c_2^{\bar \psi_{\tilde X_A}} (u, \omega, q^2)  
   & = \frac{m_B}{8 \pi^2 \left(m_B - \omega \right)^4}
   \Bigl(q^2 \left(\omega  m_B+m_B^2+4 \omega  m_c-2 \omega ^2\right) 
\nonumber  \\[1mm]
   & + m_B \bigl(m_B \left(-8
   \omega  m_c+m_c^2-3 \omega ^2\right)+m_B^2 \left(4 m_c+3 \omega \right)
\nonumber  \\[1mm]
   & - m_B^3 + 4 \omega ^2 m_c -\omega m_c^2-4 m_c^3+\omega ^3\bigr)\Bigr)\,,
 \\[4mm]
   c_3^{\bar \psi_{\tilde X_A}} (u, \omega, q^2)  
   & = -\frac{m_B^2 \left(m_B+m_c-\omega \right)}{4
   \pi^2 \left(m_B-\omega \right)^5}
   \Bigl(2 q^2 \bigl(m_B^2 \left(m_c+3 \omega \right)-3 \omega 
   m_B \left(m_c+\omega \right)
\nonumber \\[1mm]
   & - m_B^3 + \omega  \left(m_c+\omega \right)^2\bigr)+m_B
   \left(m_B-m_c-\omega \right)^3 \left(m_B+m_c-\omega \right)
\nonumber \\
   & + q^4 \left(m_B-2 \omega \right) \Bigr)\,,
\end{align}
\begin{align}
   c_2^{\bar \psi_{\tilde Y_A}} (u, \omega, q^2)  
   & = -\frac{3 m_B^2 m_c}{2 \pi ^2 \left(m_B-\omega \right){}^3} \left(m_B+m_c-\omega \right)\,,
\\[4mm]
   c_3^{\bar \psi_{\tilde Y_A}} (u, \omega, q^2)  
   & = -\frac{3 m_B^2 m_c \left(m_B+m_c-\omega \right)}{2 \pi ^2 \left(m_B-\omega \right){}^4}
   \Bigl(m_B \left(-2 \omega  m_B+m_B^2-m_c^2+\omega
   ^2\right)
\nonumber \\[1mm]
   &-q^2 \left(m_B-2 \omega \right)\Bigr)\,,
\\[4mm]
 c_2^{\bar \psi_{W}} (u, \omega, q^2)  
 & = \frac{3 m_B^2 m_c}{2 \pi ^2 \left(m_B-\omega \right)^3} \Bigl( m_B+(2 u-1) m_c-\omega \Bigr)\,,
\\[4mm]
    c_3^{\bar \psi_{W}} (u, \omega, q^2)  
    & = \frac{3 m_B^2 m_c}{2 \pi ^2 \left(m_B-\omega \right)^4}
    \Bigl(
    m_B \left(-2 \omega  m_B+m_B^2-m_c^2+\omega ^2\right)-q^2 \left(m_B-2 \omega
   \right)\Bigr)
\nonumber   \\[1mm]
   &\times \Bigl(m_B+(2 u-1) m_c-\omega \Bigr)\,,
\\[4mm]
 c_2^{\bar{\bar \psi}_{W}} (u, \omega, q^2)  
 & = \frac{3 (2 u-1) m_B^2 m_c^2}{2 \pi ^2 \left(m_B-\omega \right){}^4}\,,
  \\[4mm]
 c_3^{\bar{\bar \psi}_{W}} (u, \omega, q^2)  
 & = -\frac{3 m_B^2 m_c}{2 \pi ^2
   \left(m_B-\omega \right){}^5} \Bigl(q^2 \left(\omega  m_B-m_B^2+2 (1-2 u) \omega  m_c\right)
\nonumber   \\[1mm]
   & +m_B \left(-2 \omega 
   m_B+m_B^2-m_c^2+\omega ^2\right) \left(m_B+(2-4 u) m_c-\omega \right)\Bigr)\,,
 \\[4mm]
   c_4^{\bar{\bar \psi}_{W}} (u, \omega, q^2)  
   & = -\frac{3 m_B^3 m_c}{2 \pi ^2 \left(m_B - \omega \right){}^6}
   \Bigl(m_B \left(-2 \omega  m_B+m_B^2-m_c^2+\omega ^2\right)-q^2 \left(m_B - 2 \omega \right)\Bigr) 
\nonumber   \\[1mm]
   & \times \Bigl(q^2 \left(-m_B -2 u m_c + m_c + \omega \right) 
   + \left(-2 \omega m_B + m_B^2 - m_c^2 + \omega^2 \right) 
\nonumber  \\[1mm]
   & \quad \times \left(m_B - 2 u m_c + m_c - \omega \right) \Bigr)\,,
 \\[4mm] 
 c_2^{\bar{\bar \psi}_{Z}} (u, \omega, q^2)  
 & =\frac{3 m_B^2 m_c}{\pi ^2 \left(m_B-\omega \right){}^3}\,, 
\\[4mm]
 c_3^{\bar{\bar \psi}_{Z}} (u, \omega, q^2)  
 & = \frac{3 m_B^2 m_c}
 {\pi ^2 \left(m_B-\omega \right){}^4} 
 \Bigl(m_B \bigl((3-6 u) m_B m_c-2 \omega  m_B+m_B^2+3 (2 u-1) \omega  m_c
\nonumber \\[1mm]
 & - 4 m_c^2+\omega ^2\bigr) - q^2 \left(m_B-2 \omega \right) \Bigr)\,,
\\[4mm]
   c_4^{\bar{\bar \psi}_{Z}} (u, \omega, q^2)  
   & = -\frac{9 m_B^3 m_c^2}{\pi ^2 \left(m_B-\omega \right){}^5} 
  \Bigl(m_B \left(-2 \omega  m_B+m_B^2-m_c^2+\omega ^2\right)-q^2 \left(m_B-2 \omega \right)\Bigr) 
\nonumber   \\[1mm]
   & \times \Bigl((2 u-1) m_B+m_c-2 u \omega +\omega \Bigr)\,.
\end{align}

\bibliographystyle{JHEP}
\bibliography{References}

\providecommand{\href}[2]{#2}\begingroup\raggedright\begin{thebibliography}{10}

\bibitem{Workman:2022ynf}
{\bf Particle Data Group} Collaboration, R.~L. Workman et~al., {\it {Review of
  Particle Physics}},  {\em PTEP} {\bf 2022} (2022) 083C01.

\bibitem{Belle:2001ccu}
{\bf Belle} Collaboration, K.~Abe et~al., {\it {Observation of Cabibbo
  suppressed $B \to D^{(*)} K^-$ decays at BELLE}},  {\em Phys. Rev. Lett.}
  {\bf 87} (2001) 111801, [\href{http://arxiv.org/abs/hep-ex/0104051}{{\tt
  hep-ex/0104051}}].

\bibitem{CDF:2006hob}
{\bf CDF} Collaboration, A.~Abulencia et~al., {\it {Measurement of the Ratios
  of Branching Fractions ${\rm Br}(B^0_{(s)} \to D^-_{(s)} \pi^+ \pi^+ \pi^-) /
  {\rm Br}(B^0 \to D^- \pi^+ \pi^+ \pi^-)$ and ${\rm Br}(B^0_{(s)} \to
  D^-_{(s)} \pi^+) / {\rm Br}(B^0 \to D^- \pi^+)$}},  {\em Phys. Rev. Lett.}
  {\bf 98} (2007) 061802, [\href{http://arxiv.org/abs/hep-ex/0610045}{{\tt
  hep-ex/0610045}}].

\bibitem{Belle:2008ezn}
{\bf Belle} Collaboration, R.~Louvot et~al., {\it {Measurement of the Decay
  $B_s^0 \to D_s^- \pi^{+}$ and Evidence for $B_s^0 \to D_s^\mp K^\pm$ in $e^+
  e^-$ Annihilation at $\sqrt{s} \sim$ 10.87 GeV}},  {\em Phys. Rev. Lett.}
  {\bf 102} (2009) 021801, [\href{http://arxiv.org/abs/0809.2526}{{\tt
  arXiv:0809.2526}}].

\bibitem{LHCb:2013vfg}
{\bf LHCb} Collaboration, R.~Aaij et~al., {\it {Measurement of the
  fragmentation fraction ratio $f_{s}/f_{d}$ and its dependence on $B$ meson
  kinematics}},  {\em JHEP} {\bf 04} (2013) 001,
  [\href{http://arxiv.org/abs/1301.5286}{{\tt arXiv:1301.5286}}].

\bibitem{Belle:2021udv}
{\bf Belle} Collaboration, E.~Waheed et~al., {\it {Study of
  $\overline{B}{}^0\rightarrow D^{+}h^{-} (h=K/\pi)$ decays at Belle}},  {\em
  Phys. Rev. D} {\bf 105} (2022), no.~1 012003,
  [\href{http://arxiv.org/abs/2111.04978}{{\tt arXiv:2111.04978}}].

\bibitem{LHCb:2021qbv}
{\bf LHCb} Collaboration, R.~Aaij et~al., {\it {Precise measurement of
  the~$f_s/f_d$ ratio of fragmentation fractions and of $B^0_s$ decay branching
  fractions}},  {\em Phys. Rev. D} {\bf 104} (2021), no.~3 032005,
  [\href{http://arxiv.org/abs/2103.06810}{{\tt arXiv:2103.06810}}].

\bibitem{Buchalla:1995vs}
G.~Buchalla, A.~J. Buras, and M.~E. Lautenbacher, {\it {Weak decays beyond
  leading logarithms}},  {\em Rev. Mod. Phys.} {\bf 68} (1996) 1125--1144,
  [\href{http://arxiv.org/abs/hep-ph/9512380}{{\tt hep-ph/9512380}}].

\bibitem{Chetyrkin:1996vx}
K.~G. Chetyrkin, M.~Misiak, and M.~Munz, {\it {Weak radiative B meson decay
  beyond leading logarithms}},  {\em Phys. Lett. B} {\bf 400} (1997) 206--219,
  [\href{http://arxiv.org/abs/hep-ph/9612313}{{\tt hep-ph/9612313}}]. [Erratum:
  Phys.Lett.B 425, 414 (1998)].

\bibitem{Chetyrkin:1997gb}
K.~G. Chetyrkin, M.~Misiak, and M.~Munz, {\it {$|\Delta F| = 1$ nonleptonic
  effective Hamiltonian in a simpler scheme}},  {\em Nucl. Phys. B} {\bf 520}
  (1998) 279--297, [\href{http://arxiv.org/abs/hep-ph/9711280}{{\tt
  hep-ph/9711280}}].

\bibitem{Gorbahn:2004my}
M.~Gorbahn and U.~Haisch, {\it {Effective Hamiltonian for non-leptonic $|\Delta
  F| = 1$ decays at NNLO in QCD}},  {\em Nucl. Phys. B} {\bf 713} (2005)
  291--332, [\href{http://arxiv.org/abs/hep-ph/0411071}{{\tt hep-ph/0411071}}].

\bibitem{Blok:1992na}
B.~Blok and M.~A. Shifman, {\it {Nonfactorizable amplitudes in weak nonleptonic
  decays of heavy mesons}},  {\em Nucl. Phys. B} {\bf 389} (1993) 534--548,
  [\href{http://arxiv.org/abs/hep-ph/9205221}{{\tt hep-ph/9205221}}].

\bibitem{Shifman:1978bx}
M.~A. Shifman, A.~I. Vainshtein, and V.~I. Zakharov, {\it {QCD and Resonance
  Physics. Theoretical Foundations}},  {\em Nucl. Phys. B} {\bf 147} (1979)
  385--447.

\bibitem{Shifman:1978by}
M.~A. Shifman, A.~I. Vainshtein, and V.~I. Zakharov, {\it {QCD and Resonance
  Physics: Applications}},  {\em Nucl. Phys. B} {\bf 147} (1979) 448--518.

\bibitem{Halperin:1994hg}
I.~E. Halperin, {\it {Soft gluon suppression of 1/N(c) contributions in color
  suppressed heavy meson decays}},  {\em Phys. Lett. B} {\bf 349} (1995)
  548--554, [\href{http://arxiv.org/abs/hep-ph/9411422}{{\tt hep-ph/9411422}}].

\bibitem{Balitsky:1989ry}
I.~I. Balitsky, V.~M. Braun, and A.~V. Kolesnichenko, {\it {Radiative Decay
  $\Sigma^+ \to p \gamma$ in Quantum Chromodynamics}},  {\em Nucl. Phys. B}
  {\bf 312} (1989) 509--550.

\bibitem{Cui:2004jc}
J.-Y. Cui and Z.-H. Li, {\it {Soft nonfactorizable contribution to $\bar B^0
  \to D^0 \pi^0$}},  {\em Eur. Phys. J. C} {\bf 38} (2004) 187--194,
  [\href{http://arxiv.org/abs/hep-ph/0410029}{{\tt hep-ph/0410029}}].

\bibitem{Khodjamirian:2000mi}
A.~Khodjamirian, {\it {$B \to \pi \pi$ decay in QCD}},  {\em Nucl. Phys. B}
  {\bf 605} (2001) 558--578, [\href{http://arxiv.org/abs/hep-ph/0012271}{{\tt
  hep-ph/0012271}}].

\bibitem{Beneke:1999br}
M.~Beneke, G.~Buchalla, M.~Neubert, and C.~T. Sachrajda, {\it {QCD
  factorization for $B \to \pi \pi$ decays: Strong phases and CP violation in
  the heavy quark limit}},  {\em Phys. Rev. Lett.} {\bf 83} (1999) 1914--1917,
  [\href{http://arxiv.org/abs/hep-ph/9905312}{{\tt hep-ph/9905312}}].

\bibitem{Beneke:2000ry}
M.~Beneke, G.~Buchalla, M.~Neubert, and C.~T. Sachrajda, {\it {QCD
  factorization for exclusive, nonleptonic B meson decays: General arguments
  and the case of heavy light final states}},  {\em Nucl. Phys. B} {\bf 591}
  (2000) 313--418, [\href{http://arxiv.org/abs/hep-ph/0006124}{{\tt
  hep-ph/0006124}}].

\bibitem{Beneke:2001ev}
M.~Beneke, G.~Buchalla, M.~Neubert, and C.~T. Sachrajda, {\it {QCD
  factorization in $B \to \pi K, \pi \pi$ decays and extraction of Wolfenstein
  parameters}},  {\em Nucl. Phys. B} {\bf 606} (2001) 245--321,
  [\href{http://arxiv.org/abs/hep-ph/0104110}{{\tt hep-ph/0104110}}].

\bibitem{Huber:2016xod}
T.~Huber, S.~Kr\"ankl, and X.-Q. Li, {\it {Two-body non-leptonic heavy-to-heavy
  decays at NNLO in QCD factorization}},  {\em JHEP} {\bf 09} (2016) 112,
  [\href{http://arxiv.org/abs/1606.02888}{{\tt arXiv:1606.02888}}].

\bibitem{Bordone:2020gao}
M.~Bordone, N.~Gubernari, T.~Huber, M.~Jung, and D.~van Dyk, {\it {A puzzle in
  $\bar{B}_{(s)}^0 \to D_{(s)}^+ \lbrace \pi^-, K^-\rbrace$ decays and
  extraction of the $f_s/f_d$ fragmentation fraction}},  {\em Eur. Phys. J. C}
  {\bf 80} (2020), no.~10 951, [\href{http://arxiv.org/abs/2007.10338}{{\tt
  arXiv:2007.10338}}].

\bibitem{Bordone:2019guc}
M.~Bordone, N.~Gubernari, D.~van Dyk, and M.~Jung, {\it {Heavy-Quark expansion
  for ${{\bar{B}}_s\rightarrow D^{(*)}_s}$ form factors and unitarity bounds
  beyond the ${SU(3)_F}$ limit}},  {\em Eur. Phys. J. C} {\bf 80} (2020), no.~4
  347, [\href{http://arxiv.org/abs/1912.09335}{{\tt arXiv:1912.09335}}].

\bibitem{FermilabLattice:2014ysv}
{\bf Fermilab Lattice, MILC} Collaboration, J.~A. Bailey et~al., {\it {Update
  of $|V_{cb}|$ from the $\bar{B}\to D^*\ell\bar{\nu}$ form factor at zero
  recoil with three-flavor lattice QCD}},  {\em Phys. Rev. D} {\bf 89} (2014),
  no.~11 114504, [\href{http://arxiv.org/abs/1403.0635}{{\tt
  arXiv:1403.0635}}].

\bibitem{MILC:2015uhg}
{\bf MILC} Collaboration, J.~A. Bailey et~al., {\it
  {B\textrightarrow{}D\ensuremath{\ell}\ensuremath{\nu} form factors at nonzero
  recoil and |V$_{cb}$| from 2+1-flavor lattice QCD}},  {\em Phys. Rev. D} {\bf
  92} (2015), no.~3 034506, [\href{http://arxiv.org/abs/1503.07237}{{\tt
  arXiv:1503.07237}}].

\bibitem{Na:2015kha}
{\bf HPQCD} Collaboration, H.~Na, C.~M. Bouchard, G.~P. Lepage, C.~Monahan, and
  J.~Shigemitsu, {\it {$B \rightarrow D l \nu$ form factors at nonzero recoil
  and extraction of $|V_{cb}|$}},  {\em Phys. Rev. D} {\bf 92} (2015), no.~5
  054510, [\href{http://arxiv.org/abs/1505.03925}{{\tt arXiv:1505.03925}}].
  [Erratum: Phys.Rev.D 93, 119906 (2016)].

\bibitem{Harrison:2017fmw}
{\bf HPQCD} Collaboration, J.~Harrison, C.~Davies, and M.~Wingate, {\it
  {Lattice QCD calculation of the ${{B}_{(s)}\to D_{(s)}^{*}\ell{\nu}}$ form
  factors at zero recoil and implications for ${|V_{cb}|}$}},  {\em Phys. Rev.
  D} {\bf 97} (2018), no.~5 054502,
  [\href{http://arxiv.org/abs/1711.11013}{{\tt arXiv:1711.11013}}].

\bibitem{McLean:2019qcx}
E.~McLean, C.~T.~H. Davies, J.~Koponen, and A.~T. Lytle, {\it {$B_s\to D_s
  \ell\nu$ Form Factors for the full $q^2$ range from Lattice QCD with
  non-perturbatively normalized currents}},  {\em Phys. Rev. D} {\bf 101}
  (2020), no.~7 074513, [\href{http://arxiv.org/abs/1906.00701}{{\tt
  arXiv:1906.00701}}].

\bibitem{McLean:2019sds}
E.~McLean, C.~T.~H. Davies, A.~T. Lytle, and J.~Koponen, {\it {Lattice QCD form
  factor for $B_s\to D_s^* l\nu$ at zero recoil with non-perturbative current
  renormalisation}},  {\em Phys. Rev. D} {\bf 99} (2019), no.~11 114512,
  [\href{http://arxiv.org/abs/1904.02046}{{\tt arXiv:1904.02046}}].

\bibitem{Gubernari:2018wyi}
N.~Gubernari, A.~Kokulu, and D.~van Dyk, {\it {$B\to P$ and $B\to V$ Form
  Factors from $B$-Meson Light-Cone Sum Rules beyond Leading Twist}},  {\em
  JHEP} {\bf 01} (2019) 150, [\href{http://arxiv.org/abs/1811.00983}{{\tt
  arXiv:1811.00983}}].

\bibitem{Fleischer:2016dqd}
R.~Fleischer and K.~K. Vos, {\it {$B^0_s$-$\bar B^0_s$ Oscillations as a New
  Tool to Explore CP Violation in $D_s^\pm$ Decays}},  {\em Phys. Lett. B} {\bf
  770} (2017) 319--324, [\href{http://arxiv.org/abs/1606.06042}{{\tt
  arXiv:1606.06042}}].

\bibitem{Gershon:2021pnc}
T.~Gershon, A.~Lenz, A.~V. Rusov, and N.~Skidmore, {\it {Testing the Standard
  Model with CP asymmetries in flavor-specific nonleptonic decays}},  {\em
  Phys. Rev. D} {\bf 105} (2022), no.~11 115023,
  [\href{http://arxiv.org/abs/2111.04478}{{\tt arXiv:2111.04478}}].

\bibitem{Cai:2021mlt}
F.-M. Cai, W.-J. Deng, X.-Q. Li, and Y.-D. Yang, {\it {Probing new physics in
  class-I B-meson decays into heavy-light final states}},  {\em JHEP} {\bf 10}
  (2021) 235, [\href{http://arxiv.org/abs/2103.04138}{{\tt arXiv:2103.04138}}].

\bibitem{Beneke:2021jhp}
M.~Beneke, P.~B\"oer, G.~Finauri, and K.~K. Vos, {\it {QED factorization of
  two-body non-leptonic and semi-leptonic B to charm decays}},  {\em JHEP} {\bf
  10} (2021) 223, [\href{http://arxiv.org/abs/2107.03819}{{\tt
  arXiv:2107.03819}}].

\bibitem{Endo:2021ifc}
M.~Endo, S.~Iguro, and S.~Mishima, {\it {Revisiting rescattering contributions
  to $ \overline{B} _{(s)}$ \textrightarrow{} $ {D}_{(s)}^{\left(\ast \right)}M
  $ decays}},  {\em JHEP} {\bf 01} (2022) 147,
  [\href{http://arxiv.org/abs/2109.10811}{{\tt arXiv:2109.10811}}].

\bibitem{Iguro:2020ndk}
S.~Iguro and T.~Kitahara, {\it {Implications for new physics from a novel
  puzzle in $\bar{B}_{(s)}^0 \to D^{(\ast)+}_{(s)} \lbrace \pi^-, K^- \rbrace$
  decays}},  {\em Phys. Rev. D} {\bf 102} (2020), no.~7 071701,
  [\href{http://arxiv.org/abs/2008.01086}{{\tt arXiv:2008.01086}}].

\bibitem{Bordone:2021cca}
M.~Bordone, A.~Greljo, and D.~Marzocca, {\it {Exploiting dijet resonance
  searches for flavor physics}},  {\em JHEP} {\bf 08} (2021) 036,
  [\href{http://arxiv.org/abs/2103.10332}{{\tt arXiv:2103.10332}}].

\bibitem{Fleischer:2021cct}
R.~Fleischer and E.~Malami, {\it {Using $B^0_s\to D_s^\mp K^\pm$ Decays as a
  Portal to New Physics}},  {\em Phys. Rev. D} {\bf 106} (2022), no.~5 056004,
  [\href{http://arxiv.org/abs/2109.04950}{{\tt arXiv:2109.04950}}].

\bibitem{Fleischer:2021cwb}
R.~Fleischer and E.~Malami, {\it {Revealing new physics in
  ${\bar{B}}^{0}_{s}\rightarrow D_s^{\mp } K^{\pm }$ decays}},  {\em Eur. Phys.
  J. C} {\bf 83} (2023), no.~5 420,
  [\href{http://arxiv.org/abs/2110.04240}{{\tt arXiv:2110.04240}}].

\bibitem{Lenz:2022pgw}
A.~Lenz, J.~M\"uller, M.~L. Piscopo, and A.~V. Rusov, {\it {Taming New Physics
  in $b \to c \bar u d (s) $ with $\tau (B^+)/\tau(B_d)$ and $a_{sl}^d$}},
  \href{http://arxiv.org/abs/2211.02724}{{\tt arXiv:2211.02724}}.

\bibitem{Braun:2017liq}
V.~M. Braun, Y.~Ji, and A.~N. Manashov, {\it {Higher-twist B-meson Distribution
  Amplitudes in HQET}},  {\em JHEP} {\bf 05} (2017) 022,
  [\href{http://arxiv.org/abs/1703.02446}{{\tt arXiv:1703.02446}}].

\bibitem{Buras:1989xd}
A.~J. Buras and P.~H. Weisz, {\it {QCD Nonleading Corrections to Weak Decays in
  Dimensional Regularization and 't Hooft-Veltman Schemes}},  {\em Nucl. Phys.
  B} {\bf 333} (1990) 66--99.

\bibitem{Novikov:1983gd}
V.~A. Novikov, M.~A. Shifman, A.~I. Vainshtein, and V.~I. Zakharov, {\it
  {Calculations in External Fields in Quantum Chromodynamics. Technical
  Review}},  {\em Fortsch. Phys.} {\bf 32} (1984) 585.

\bibitem{Balitsky:1987bk}
I.~I. Balitsky and V.~M. Braun, {\it {Evolution Equations for QCD String
  Operators}},  {\em Nucl. Phys. B} {\bf 311} (1989) 541--584.

\bibitem{Belyaev:1994zk}
V.~M. Belyaev, V.~M. Braun, A.~Khodjamirian, and R.~Ruckl, {\it {D* D pi and B*
  B pi couplings in QCD}},  {\em Phys. Rev. D} {\bf 51} (1995) 6177--6195,
  [\href{http://arxiv.org/abs/hep-ph/9410280}{{\tt hep-ph/9410280}}].

\bibitem{Shtabovenko:2020gxv}
V.~Shtabovenko, R.~Mertig, and F.~Orellana, {\it {FeynCalc 9.3: New features
  and improvements}},  {\em Comput. Phys. Commun.} {\bf 256} (2020) 107478,
  [\href{http://arxiv.org/abs/2001.04407}{{\tt arXiv:2001.04407}}].

\bibitem{Khodjamirian:2010vf}
A.~Khodjamirian, T.~Mannel, A.~A. Pivovarov, and Y.~M. Wang, {\it {Charm-loop
  effect in $B \to K^{(*)} \ell^{+} \ell^{-}$ and $B\to K^*\gamma$}},  {\em
  JHEP} {\bf 09} (2010) 089, [\href{http://arxiv.org/abs/1006.4945}{{\tt
  arXiv:1006.4945}}].

\bibitem{Gubernari:2020eft}
N.~Gubernari, D.~van Dyk, and J.~Virto, {\it {Non-local matrix elements in
  $B_{(s)}\to \{K^{(*)},\phi\}\ell^+\ell^-$}},  {\em JHEP} {\bf 02} (2021) 088,
  [\href{http://arxiv.org/abs/2011.09813}{{\tt arXiv:2011.09813}}].

\bibitem{Geyer:2005fb}
B.~Geyer and O.~Witzel, {\it {B-meson distribution amplitudes of geometric
  twist vs. dynamical twist}},  {\em Phys. Rev. D} {\bf 72} (2005) 034023,
  [\href{http://arxiv.org/abs/hep-ph/0502239}{{\tt hep-ph/0502239}}].

\bibitem{Albrecht:2021tul}
J.~Albrecht, D.~van Dyk, and C.~Langenbruch, {\it {Flavour anomalies in heavy
  quark decays}},  {\em Prog. Part. Nucl. Phys.} {\bf 120} (2021) 103885,
  [\href{http://arxiv.org/abs/2107.04822}{{\tt arXiv:2107.04822}}].

\bibitem{London:2021lfn}
D.~London and J.~Matias, {\it {$B$ Flavour Anomalies: 2021 Theoretical Status
  Report}},  {\em Ann. Rev. Nucl. Part. Sci.} {\bf 72} (2022) 37--68,
  [\href{http://arxiv.org/abs/2110.13270}{{\tt arXiv:2110.13270}}].

\bibitem{Colangelo:2000dp}
P.~Colangelo and A.~Khodjamirian, {\it {QCD sum rules, a modern perspective}},
  \href{http://arxiv.org/abs/hep-ph/0010175}{{\tt hep-ph/0010175}}.

\bibitem{Khodjamirian:2020btr}
A.~Khodjamirian, {\em {Hadron Form Factors}: {From Basic Phenomenology to QCD
  Sum Rules}}.
\newblock CRC Press, Taylor \& Francis Group, Boca Raton, FL, USA, 2020.

\bibitem{Iagolnitzer:1991wj}
D.~Iagolnitzer, {\it {Causality in local quantum field theory: Some general
  results}},  {\em Commun. Math. Phys.} {\bf 144} (1992) 235--256.

\bibitem{Muta:2010xua}
T.~Muta, {\it {Foundations of Quantum Chromodynamics: An Introduction to
  Perturbative Methods in Gauge Theories, (3rd ed.)}},  {\em World Scientific,
  Hackensack, N.J., 3rd ed.,} (2010).

\bibitem{Ruiz:2023ozv}
R.~Ruiz et~al., {\it {Target mass corrections in lepton-nucleus DIS: theory and
  applications to nuclear PDFs}},  \href{http://arxiv.org/abs/2301.07715}{{\tt
  arXiv:2301.07715}}.

\bibitem{Benzke:2010js}
M.~Benzke, S.~J. Lee, M.~Neubert, and G.~Paz, {\it {Factorization at Subleading
  Power and Irreducible Uncertainties in $\bar B\to X_s\gamma$ Decay}},  {\em
  JHEP} {\bf 08} (2010) 099, [\href{http://arxiv.org/abs/1003.5012}{{\tt
  arXiv:1003.5012}}].

\bibitem{Bell:2013tfa}
G.~Bell, T.~Feldmann, Y.-M. Wang, and M.~W.~Y. Yip, {\it {Light-Cone
  Distribution Amplitudes for Heavy-Quark Hadrons}},  {\em JHEP} {\bf 11}
  (2013) 191, [\href{http://arxiv.org/abs/1308.6114}{{\tt arXiv:1308.6114}}].

\bibitem{Qin:2022rlk}
Q.~Qin, Y.-L. Shen, C.~Wang, and Y.-M. Wang, {\it {Deciphering the
  long-distance penguin contribution to $\bar B_{d, s} \to \gamma \gamma$
  decays}},  \href{http://arxiv.org/abs/2207.02691}{{\tt arXiv:2207.02691}}.

\bibitem{Melikhov:2022wct}
D.~Melikhov, {\it {Nonfactorizable charming loops in FCNC B decays versus
  B-decay semileptonic form factors}},  {\em Phys. Rev. D} {\bf 106} (2022),
  no.~5 054022, [\href{http://arxiv.org/abs/2208.04907}{{\tt
  arXiv:2208.04907}}].

\bibitem{Melikhov:2023pet}
D.~Melikhov, {\it {Three-particle distribution in B meson and charm-quark loops
  in FCNC B decays}},  \href{http://arxiv.org/abs/2302.13673}{{\tt
  arXiv:2302.13673}}.

\bibitem{Shifman:2000jv}
M.~A. Shifman, {\it {Quark hadron duality}},  in {\em {8th International
  Symposium on Heavy Flavor Physics}}, vol.~3, (Singapore), pp.~1447--1494,
  World Scientific, 7, 2000.
\newblock \href{http://arxiv.org/abs/hep-ph/0009131}{{\tt hep-ph/0009131}}.

\bibitem{Rusov:2017chr}
A.~V. Rusov, {\it {Higher-twist effects in light-cone sum rule for the
  $B\rightarrow \pi $ form factor}},  {\em Eur. Phys. J. C} {\bf 77} (2017),
  no.~7 442, [\href{http://arxiv.org/abs/1705.01929}{{\tt arXiv:1705.01929}}].

\bibitem{Khodjamirian:2006st}
A.~Khodjamirian, T.~Mannel, and N.~Offen, {\it {Form-factors from light-cone
  sum rules with B-meson distribution amplitudes}},  {\em Phys. Rev. D} {\bf
  75} (2007) 054013, [\href{http://arxiv.org/abs/hep-ph/0611193}{{\tt
  hep-ph/0611193}}].

\bibitem{Lu:2018cfc}
C.-D. L\"u, Y.-L. Shen, Y.-M. Wang, and Y.-B. Wei, {\it {QCD calculations of $B
  \to \pi, K$ form factors with higher-twist corrections}},  {\em JHEP} {\bf
  01} (2019) 024, [\href{http://arxiv.org/abs/1810.00819}{{\tt
  arXiv:1810.00819}}].

\bibitem{Beneke:2018wjp}
M.~Beneke, V.~M. Braun, Y.~Ji, and Y.-B. Wei, {\it {Radiative leptonic decay
  $B\to \gamma \ell \nu_\ell$ with subleading power corrections}},  {\em JHEP}
  {\bf 07} (2018) 154, [\href{http://arxiv.org/abs/1804.04962}{{\tt
  arXiv:1804.04962}}].

\bibitem{Feldmann:2022uok}
T.~Feldmann, P.~L\"ughausen, and D.~van Dyk, {\it {Systematic parametrization
  of the leading B-meson light-cone distribution amplitude}},  {\em JHEP} {\bf
  10} (2022) 162, [\href{http://arxiv.org/abs/2203.15679}{{\tt
  arXiv:2203.15679}}].

\bibitem{Braun:2003wx}
V.~M. Braun, D.~Y. Ivanov, and G.~P. Korchemsky, {\it {The B meson distribution
  amplitude in QCD}},  {\em Phys. Rev. D} {\bf 69} (2004) 034014,
  [\href{http://arxiv.org/abs/hep-ph/0309330}{{\tt hep-ph/0309330}}].

\bibitem{Khodjamirian:2020hob}
A.~Khodjamirian, R.~Mandal, and T.~Mannel, {\it {Inverse moment of the
  B$_{s}$-meson distribution amplitude from QCD sum rule}},  {\em JHEP} {\bf
  10} (2020) 043, [\href{http://arxiv.org/abs/2008.03935}{{\tt
  arXiv:2008.03935}}].

\bibitem{Lee:2005gza}
S.~J. Lee and M.~Neubert, {\it {Model-independent properties of the B-meson
  distribution amplitude}},  {\em Phys. Rev. D} {\bf 72} (2005) 094028,
  [\href{http://arxiv.org/abs/hep-ph/0509350}{{\tt hep-ph/0509350}}].

\bibitem{Kawamura:2008vq}
H.~Kawamura and K.~Tanaka, {\it {Operator product expansion for B-meson
  distribution amplitude and dimension-5 HQET operators}},  {\em Phys. Lett. B}
  {\bf 673} (2009) 201--207, [\href{http://arxiv.org/abs/0810.5628}{{\tt
  arXiv:0810.5628}}].

\bibitem{Kawamura:2010tj}
H.~Kawamura and K.~Tanaka, {\it {Evolution equation for the B-meson
  distribution amplitude in the heavy-quark effective theory in coordinate
  space}},  {\em Phys. Rev. D} {\bf 81} (2010) 114009,
  [\href{http://arxiv.org/abs/1002.1177}{{\tt arXiv:1002.1177}}].

\bibitem{Feldmann:2023aml}
T.~Feldmann, P.~L\"ughausen, and N.~Seitz, {\it {Strange-quark mass effects in
  the $B_s$ meson's light-cone distribution amplitude}},
  \href{http://arxiv.org/abs/2306.14686}{{\tt arXiv:2306.14686}}.

\bibitem{Ball:2003fq}
P.~Ball and E.~Kou, {\it {$B \to \gamma e \nu$ transitions from QCD sum rules
  on the light cone}},  {\em JHEP} {\bf 04} (2003) 029,
  [\href{http://arxiv.org/abs/hep-ph/0301135}{{\tt hep-ph/0301135}}].

\bibitem{Beneke:2011nf}
M.~Beneke and J.~Rohrwild, {\it {B meson distribution amplitude from $B \to
  \gamma \ell \nu$}},  {\em Eur. Phys. J. C} {\bf 71} (2011) 1818,
  [\href{http://arxiv.org/abs/1110.3228}{{\tt arXiv:1110.3228}}].

\bibitem{Braun:2012kp}
V.~M. Braun and A.~Khodjamirian, {\it {Soft contribution to $B\to \gamma \ell
  \nu_\ell$ and the $B$-meson distribution amplitude}},  {\em Phys. Lett. B}
  {\bf 718} (2013) 1014--1019, [\href{http://arxiv.org/abs/1210.4453}{{\tt
  arXiv:1210.4453}}].

\bibitem{Grozin:1996pq}
A.~G. Grozin and M.~Neubert, {\it {Asymptotics of heavy meson form-factors}},
  {\em Phys. Rev. D} {\bf 55} (1997) 272--290,
  [\href{http://arxiv.org/abs/hep-ph/9607366}{{\tt hep-ph/9607366}}].

\bibitem{Nishikawa:2011qk}
T.~Nishikawa and K.~Tanaka, {\it {QCD Sum Rules for Quark-Gluon Three-Body
  Components in the B Meson}},  {\em Nucl. Phys. B} {\bf 879} (2014) 110--142,
  [\href{http://arxiv.org/abs/1109.6786}{{\tt arXiv:1109.6786}}].

\bibitem{Rahimi:2020zzo}
M.~Rahimi and M.~Wald, {\it {QCD sum rules for parameters of the B-meson
  distribution amplitudes}},  {\em Phys. Rev. D} {\bf 104} (2021), no.~1
  016027, [\href{http://arxiv.org/abs/2012.12165}{{\tt arXiv:2012.12165}}].

\bibitem{Khodjamirian:2005ea}
A.~Khodjamirian, T.~Mannel, and N.~Offen, {\it {B-meson distribution amplitude
  from the $B \to \pi$ form-factor}},  {\em Phys. Lett. B} {\bf 620} (2005)
  52--60, [\href{http://arxiv.org/abs/hep-ph/0504091}{{\tt hep-ph/0504091}}].

\bibitem{Faller:2008tr}
S.~Faller, A.~Khodjamirian, C.~Klein, and T.~Mannel, {\it {$B \to D^{(*)}$ Form
  Factors from QCD Light-Cone Sum Rules}},  {\em Eur. Phys. J. C} {\bf 60}
  (2009) 603--615, [\href{http://arxiv.org/abs/0809.0222}{{\tt
  arXiv:0809.0222}}].

\bibitem{Khodjamirian:2003xk}
A.~Khodjamirian, T.~Mannel, and M.~Melcher, {\it {Flavor SU(3) symmetry in
  charmless B decays}},  {\em Phys. Rev. D} {\bf 68} (2003) 114007,
  [\href{http://arxiv.org/abs/hep-ph/0308297}{{\tt hep-ph/0308297}}].

\bibitem{FLAG:2019iem}
{\bf Flavour Lattice Averaging Group} Collaboration, S.~Aoki et~al., {\it {FLAG
  Review 2019: Flavour Lattice Averaging Group (FLAG)}},  {\em Eur. Phys. J. C}
  {\bf 80} (2020), no.~2 113, [\href{http://arxiv.org/abs/1902.08191}{{\tt
  arXiv:1902.08191}}].

\bibitem{Neubert:1991sp}
M.~Neubert, {\it {Heavy meson form-factors from QCD sum rules}},  {\em Phys.
  Rev. D} {\bf 45} (1992) 2451--2466.

\bibitem{Grozin:1996hk}
A.~G. Grozin and M.~Neubert, {\it {Hybrid renormalization of penguins and
  five-dimension heavy light operators}},  {\em Nucl. Phys. B} {\bf 495} (1997)
  81--98, [\href{http://arxiv.org/abs/hep-ph/9701262}{{\tt hep-ph/9701262}}].

\bibitem{Herren:2017osy}
F.~Herren and M.~Steinhauser, {\it {Version 3 of RunDec and CRunDec}},  {\em
  Comput. Phys. Commun.} {\bf 224} (2018) 333--345,
  [\href{http://arxiv.org/abs/1703.03751}{{\tt arXiv:1703.03751}}].

\bibitem{Lenz:2022rbq}
A.~Lenz, M.~L. Piscopo, and A.~V. Rusov, {\it {Disintegration of beauty: a
  precision study}},  {\em JHEP} {\bf 01} (2023) 004,
  [\href{http://arxiv.org/abs/2208.02643}{{\tt arXiv:2208.02643}}].

\bibitem{Neubert:1993mb}
M.~Neubert, {\it {Heavy quark symmetry}},  {\em Phys. Rept.} {\bf 245} (1994)
  259--396, [\href{http://arxiv.org/abs/hep-ph/9306320}{{\tt hep-ph/9306320}}].

\end{thebibliography}\endgroup

\end{document}